\documentclass[aps,prd,preprintnumbers,nofootinbib,twocolumn,longbibliography]{revtex4-1}
\usepackage{hyperref}
\usepackage{amsfonts}
\usepackage{amsmath}

\usepackage{amssymb}
\usepackage{graphicx}%
\usepackage{bigints}
\usepackage{graphicx}
\usepackage{subfigure}
\usepackage{epsf}
\usepackage{bm}
\usepackage{amsmath}
\usepackage{amsfonts}
\usepackage{amssymb}
\usepackage{graphicx}
\usepackage{tabularx}
\usepackage{color}%
\providecommand{\U}[1]{\protect\rule{.1in}{.1in}}

\newcommand{\be}{\begin{equation}}
\newcommand{\ee}{\end{equation}}

\newcommand{\mincir}{\raise
-3.truept\hbox{\rlap{\hbox{$\sim$}}\raise4.truept\hbox{$<$}\ }}
\newcommand{\magcir}{\raise
-3.truept\hbox{\rlap{\hbox{$\sim$}}\raise4.truept\hbox{$>$}\ }}

\usepackage{physics}
\providecommand{\U}[1]{\protect\rule{.1in}{.1in}}

\usepackage{tikz,xcolor,hyperref}

\begin{document}

\title{Generalized Uncertainty Principle from the Regularized Self-Energy}
\author{Kimet Jusufi$^\odot$}
\email{kimet.jusufi@unite.edu.mk}

\author{Ahmed~Farag Ali$^{\triangle \nabla}$}
\email{aali29@essex.edu ;  ahmed.ali@fsc.bu.edu.eg}

\affiliation{$^\odot$ Physics Department, State University of Tetovo, Ilinden Street nn, 1200, Tetovo, North Macedonia}

\affiliation{$^\triangle$Essex County College, 303 University Ave, Newark, NJ 07102, United States.}

\affiliation{$^\nabla$Department of Physics, Faculty of Science, Benha University, Benha, 13518, Egypt.}

\begin{abstract}
We use the Schr\"odinger--Newton equation to calculate the regularized self-energy of the particle using a regular self-gravitational and electrostatic potential derived in the string T-duality. The particle mass $M$ is no longer concentrated into a point but it is diluted and described by a quantum-corrected smeared energy density resulting in corrections to the energy of the particle which is interpreted as a regularized self-energy. We extend our results and find corrections to the relativistic particles using the Klein-Gordon, Proca, and Dirac equations. An important finding is that we extract a form of generalized uncertainty principle (GUP) from the corrected energy. The form of GUP is shown to depend on the nature of particles; namely, for bosons (spin $0$ and spin $1$) we obtain a quadratic form of GUP, while for fermions (spin $1/2$) we obtain a linear form of GUP. The correlation we found between spin and GUP may offer insights into investigating quantum gravity.

. 
\end{abstract}

\maketitle
\section{Introduction}

A significant challenge in the field of physics remains unresolved, which is how to integrate the principles of quantum mechanics and gravity into a quantum theory of gravity.
The fundamental equation of gravity is Einstein's field equation \cite{Einstein:1916vd}.  The fundamental equation in quantum mechanics is Schrödinger's equation  \cite{Schrodinger:1926gei}, while Klein-Gordon equation \cite{gordon1926comptoneffekt,klein1926quantentheorie}, and Dirac equation \cite{Dirac:1928hu}, are viewed as relativistic field equations. A modification to Schrödinger's equation was suggested by Di{\'o}si and Penrose \cite{Diosi:1984wuz,Diosi:1986nu,Penrose:1996cv,Diosi:1988uy,penrose1998quantum,Penrose:2014nha} to include the force of gravity. It is known as the Schrödinger-Newton equation. This new term takes into account the gravitational potential generated by the distribution of mass in the system. The effect can be generalized for the Dirac equation, Klein-Gordon equation, and Proca equation \cite{Proca:1936fbw}.


In this work, we would like to study a specific form of gravitational potential that arises from string T-duality \cite{Sathiapalan:1986zb,Strominger:1996it} with these equations. On the other hand, a new spin forward was made by Padmanabhan who related the concept of duality and zero point length as a tool to obtain quantum gravity effect \cite{Padmanabhan:1996ap} and was applied to black hole physics to cure the black hole singularity for regular and charged solutions \cite{Nicolini:2019irw, Gaete:2022ukm, Nicolini:2022rlz}. In Ref. \cite{ Jusufi:2023pzt} one of the authors of the present work argued how the zero-point length which modifies the black hole spacetime can lead to a non-singular solution during the gravitational collapse. In the context of cosmology, however, it was found a modified Friedmann's equation \cite{Jusufi:2022mir} and in particular, it was found how the zero-point length corrections can lead to a bouncing scenario (see, \cite{Jusufi:2022mir,Millano:2023ahb}).  Such a potential is regular at $r=0$ and modifies the energy of the system. In this picture, the particle is no longer concentrated into a point but it is diluted along with smeared energy density. The gravitational potential implied by T-duality suggests a modification of energy that agrees with the modifications implied by generalized uncertainty principle models \cite{Amati:1988tn,Ali:2009zq,Kempf:1994su,Brau:1999uv,Maggiore:1993rv,Adler:1993hm,Scardigli:1999jh}, see also \cite{Carr:2015nqa,Carr:2022ndy}. We realize a new connection between the spin concept of particles and these energy modifications. 


Phenomenological and experimental implications of the GUP have been investigated in low and high-energy regimes. These include atomic systems \cite{Ali:2011fa, Das:2008kaa}, quantum optical systems \cite{Pikovski:2011zk}, gravitational bar detectors \cite{Marin:2013pga}, gravitational decoherence \cite{Petruzziello:2020wkd}, gravitational waves \cite{Feng:2016tyt,das2021bounds,Moussa:2021qlz},  gravitational tests \cite{Scardigli:2014qka,Hemeda:2022dnd}, composite particles \cite{Kumar:2019bnd}, astrophysical systems \cite{Moradpour:2019wpj}, condensed matter systems \cite{Iorio:2017vtw}, cold atoms \cite{gao2016constraining}, macroscopic harmonic oscillators \cite{Bawaj:2014cda}, gauge theories \cite{Kober:2010sj}, quantum noise \cite{Girdhar:2020kfl}, extended GUP from deformed algebra \cite{Segreto:2022xyv}, modified gravity theories \cite{Ali:2024tbd}, cosmological models \cite{Ashoorioon:2004vm,Easther:2001fz,dabrowski2019extended,Ali:2014hma,Easther:2001fi,Ali:2015ola} and Holography \cite{Scardigli:2003kr}. Reviews of the GUP, its phenomenology, and its experimental implications can be found in \cite{Addazi:2021xuf,Hossenfelder:2012jw}. Our work establishes a connection between T-duality in string theory and the generalized uncertainty principle. Among other things, in this work, we would like to understand if the form of GUP depends on the nature of particles, for example, the spin. It is possible to interpret that the experimental and phenomenological research on GUP provides support in favor of T-duality and string theory.

The paper is organized as follows. In Sec. II, we study the modified Schr\"odinger-Newton equation to compute the corrections to the energy of the particle. In Sec. III, we compute the self-regularized energy due to the electric field. In Sec. IV, we study the modified Klein-Gordon equation and the relativistic energy corrections. In Sec. V, we investigate the modified Dirac equation and its energy corrections. In Sec. VI, we study the modified Proca equation. In Sec. VII, we show a link between the GUP and self-regularized energy. Importantly, here we shall elaborate on how the nature of particles, say the spin of the particle, can lead to a different form of GUP. Finally in Sec. VIII, we comment on our results.

\section{Modified Schr\"odinger--Newton equation}\label{Sec_GUP}
We shall review the quantum-corrected static interaction potential according to the field theory enjoying the path integral duality. We recall that the momentum space propagator induced by the path integral duality for the massless case is given by \cite{Nicolini:2019irw}
\begin{equation}
G(k)
= -\frac{l_0}{\sqrt{k^2}}\, K_1 \!\left(l_0 \sqrt{k^2}\right)
\end{equation}
with $l_0$ being the zero-point length and $K_1\!\left(x\right)$ is a modified Bessel function of the second kind. One has two cases: at small momenta, \textit{i.e.}~for ${(l_0 k)}^2 \to 0$, we obtain the conventional massless propagator $G(k) = -k^{-2}$, at large momenta, interestingly, the exponential suppression is responsible for curing UV divergences \cite{Nicolini:2019irw}
\begin{equation}
G(k) \sim -l_0^{1/2}\, {\left(k^2\right)}^{-3/4}\, \mathrm{e}^{-l_0 \sqrt{k^2}}.
\end{equation}
Consider a static external source $J$ which consists of two point-like masses, $m$ and $M$, at relative distance $\vec{r}$, then for the potential it was found \cite{Nicolini:2019irw}
\begin{align}
V(r)
&= -\frac{1}{m}\, \frac{W[J]}{T}\\
&= -M\,G\, \int\!\frac{d^3 k}{{\left(2\pi\right)}^3}\; { G_{F}(k)|}_{k^0=0}\; 
 \exp\!\left(i \vec{k} \vec{r}\right)\\
&= -\frac{GM}{\sqrt{r^2 + l_0^2}},
\end{align}
In this section, we will use the Schr\"odinger--Newton equation to obtain the modified energy-time uncertainty relation.  The Schr\"odinger--Newton equation for a free particle with the self-gravitational potential $V$ produced by a quantum source in state $\Psi$ can be written as  \cite{Diosi:1984wuz,Diosi:1986nu,Penrose:1996cv,Diosi:1988uy,penrose1998quantum,Penrose:2014nha}
\begin{equation}
i \hbar \frac{\partial \Psi(t,\vec r) }{\partial t}= \left( -\frac{\hbar^2}{2 M} \nabla^2-M^2 G \int \frac{|\Psi(t,\vec r')|^2 d^3\vec r' }{|\vec r- \vec r'| }
\right) \Psi(t,\vec r) \,, \label{SNEI}
\end{equation}
studied by Di{\'o}si and Penrose \cite{Diosi:1984wuz,Diosi:1986nu,Penrose:1996cv,Diosi:1988uy,penrose1998quantum,Penrose:2014nha}.
Along with the semiclassical Poisson equation 
\begin{equation}
 \nabla^2V=4\pi G M |\Psi|^2  
\end{equation}
 The mass density of the source is related to the quantum probability density via 
 \begin{eqnarray}
     \rho(r,t)=M |\Psi(t,\vec r)|^2. 
 \end{eqnarray}
 Our aim is to use a well-behaved expression without a singularity at $r=0$. Such regularised expression was found in the string T-duality with the gravitational potential given by \cite{Nicolini:2019irw}
\begin{align}
V(r) \to -\frac{G  M}{\sqrt{r^2 + l_0^2}}, \label{TV}
\end{align}
where $l_0$ is the zero point length. The modified potential suggests a modification to the Schr\"odinger--Newton as follows 
\begin{equation}
i \hbar \frac{\partial \Psi(t,\vec r) }{\partial t}= \left( -\frac{\hbar^2}{2 M} \nabla^2-M^2 G \int \frac{|\Psi(t,\vec r')|^2 d^3\vec r' }{\sqrt{|\vec r-\vec r'|^2 +l_0^2}}
\right) \Psi(t,\vec r). \label{SNEII}
\end{equation}


On one hand, we see that when $l_0=0$, the standard relation is obtained. On the other hand, we can use a coupling of gravity to matter by means of the semi-classical Einstein equations \cite{Bahrami:2014gwa}
\begin{equation}
\label{eqn:sce}
R_{\mu \nu} + \frac{1}{2} g_{\mu \nu} R =\frac{ 8 \pi G}{c^4}  \,
\bra{\Psi} T_{\mu \nu} \ket{\Psi}.
\end{equation}
In the Newtonian limit, the
components of the energy-momentum tensor become the Poisson equation \cite{Bahrami:2014gwa}
\begin{equation}
\label{eqn:poisson}
 \nabla^2 V = \frac{ 4 \pi G}{c^2} \bra{\Psi} \hat{T}_{00} \ket{\Psi}.
\end{equation}

In the Newtonian limit, where $T_{00}$ is the dominant term of the energy-momentum tensor, the
interaction Hamiltonian then becomes \cite{Bahrami:2014gwa}
\begin{eqnarray}
\label{eqn:hint}
 H_\text{int} &=& \int \mathrm{d}^3r \, V \,T^{00} \\
 &=& -M G \int_{0}^{2 \pi}\int_{0}^{\pi}\int_0^{\infty} \mathrm{d}^3 r \frac{\rho(\vec r)}{\sqrt{r^2+l_0^2}} \,,
\end{eqnarray}
where by using $V(r)$ and Poisson equation one can obtain \cite{Nicolini:2019irw}
\begin{equation}
\rho(r)= \frac{1}{4\pi G}\nabla^2 V(r)=\frac{3 l_0^2 M}{4 \pi \left( r^2+l_0^2\right)^{5/2}}.
\end{equation}
Using spherical coordinates 
\begin{equation}
    \mathrm{d}^3r = r^2 \sin \theta\, \mathrm{d}\theta\, \mathrm{d}\phi\, \mathrm{d}r,
\end{equation}
and integrating this equation we obtain a finite contribution
\begin{equation}
\label{eqn:hint_spher}
 H_\text{int} = - \frac{3 \pi G M^2}{16\,l_0}.
\end{equation}


Therefore the apparently nonlinear form of the Schr\"odinger--Newton equation reduces to the linear form 
\begin{equation}
\mathrm{i} \hbar \frac{\partial }{\partial t} \Psi(t,\vec r) = -\frac{\hbar^2}{2 M} \nabla^2 \Psi(t,\vec r) + \hat{H}_{\text{int}} \Psi(t,\vec r),
\end{equation}
or in terms of operators
\begin{eqnarray}
\hat{H}\Psi(t,r)&=&\left(-\frac{\hbar^2}{2 M} \nabla^2+\hat{H}_{int}\right)\Psi(t,r),
\end{eqnarray}
and if we define $ \hat{H}_{tot}=\hat{H}-\hat{H}_{int}$, we can write
\begin{eqnarray}
    \hat{H}_{tot}\Psi(t,r)&=&-\frac{\hbar^2}{2 M} \nabla^2\Psi(t,r),
\end{eqnarray}
yielding
\begin{eqnarray}
E_{tot}\Psi(t,r)&=& \left(E+ \frac{3 \pi G M^2}{16\,l_0} \right) \Psi(t,r),
\end{eqnarray}
here $E $ can be interpreted as the bare energy of the free particle and the wave function of the particle is now
\begin{equation}
    \Psi=A \exp(-\frac{i}{\hbar}(E_{tot} t-\vec p\cdot\vec r)),
\end{equation}
meaning that the total energy in the weave function is modified. 

\subsection{Significance of effects in the SN equation}
The Schr\"odinger-Newton equation reduces to $ E_{tot}\Psi(t,r)=\frac{p^2}{2M} \Psi(t,r)$. Consider a free particle solution described by the initial Gaussian wave function
\begin{equation}
\Psi(r,t) =\left( \pi \, a^2\right)^{-3/4} \, \left(1 + \frac{\mathrm{i} \, \hbar \, t}{M\, a^2}\right)^{-3/2} \, \mathrm{e}^{- \frac{r^2}{2 a^2 \left(1 + \frac{i \, \hbar \, t}{M\, a^2}\right) } },
\end{equation}
and note that the peak of $r^2 \, \Psi^*(r,t) \, \Psi(r,t)$ is located at
\begin{equation}
r_p(t) = \sqrt{a^2 + \left(\frac{\hbar \, t}{a \, M}\right)^2}.
\end{equation}
Now let us set the acceleration
\begin{equation}
    \ddot r_p(0) = \frac{\hbar^2}{a^3 \, M^2},
\end{equation}
while the peak probability equal to the acceleration due to Newtonian gravity
\begin{equation}\label{masstimate}
\ddot r_p(0) =-\frac{\mathrm{d}}{\mathrm{d}r} \, V(r_p(0))=\frac{M\, a}{(l_0^2+a^2)^{3/2}}.
\end{equation}  
From the last two equations, we get the mass 
\begin{eqnarray}
    M^3 \,a=\frac{\hbar^2}{G} \left( 1+\frac{l_0^2}{a^2} \right)^{3/2}.
\end{eqnarray}
From the last equation, we get the standard result when $l_0=0$, i.e., $M^3 a=\hbar^2/G$. 
This equation allows us to determine a critical width for a given mass value and conversely. In this way, we can estimate the regime where the effects of the Schr\"odinger-Newton are important. 


\subsection{Curved spacetime}
We can consider further corrections by considering a curved spacetime background. In particular, the general solution in the case of a static, spherically symmetric source reads
\begin{equation}
\label{eq:lineElem}
\mathrm{d}s^2 = g_{00} \, \mathrm{d}t^2 + g_{rr} \, \mathrm{d}r^2 + r^2(\mathrm{d}\theta^2 + \sin^2\theta \, \mathrm{d}\phi^2),
\end{equation}

with $g_{00}(r)= -f(r)$, and
\begin{equation}
\begin{split}
f(r)=1 - \frac{2 G m(r)}{r}=1 + 2 \Phi(r),
\end{split}
\end{equation}
where $\Phi(r)=-Gm(r)/r$. Using the energy density (15) it was found that 
\begin{equation}
\label{eq:massFct}
m(r)
= \frac{M r^3}{{\left(r^2 +l_0^2\right)}^{3/2}},
\end{equation}
along with 
\begin{equation}
\label{eq:g00}
\begin{split}
f(r)= 1 - \frac{2 G M r^2}{{\left(r^2 +l_0^2\right)}^{3/2}}.
\end{split}
\end{equation}
In this case, the Schr\"odinger--Newton equation will be
\begin{equation}
i \hbar \frac{\partial \Psi(t,\vec r) }{\partial t}= \left( -\frac{\hbar^2}{2 M} \nabla_{LB}^2+H_{int} \right)\Psi(t,\vec r).
\end{equation}
where $\nabla^2_{LB}=g^{-1/2}\partial_{\mu}(g^{\mu \nu} g^{1/2} \partial_{\nu})$ is the Laplace-Beltrami operator in the curved spacetime geometry, with $g=\det g_{\mu \nu}$. 
For the self-gravitational interaction, we get 
\begin{eqnarray}
\label{eqn:hint}\notag
 H_\text{int}&=&\int_{0}^{2 \pi}\int_{0}^{\pi}\int_0^{\infty} \Phi(r) \rho(r) \mathrm{d}^3 r \\
 &=& - \frac{3 \pi G M^2}{32 l_0}. 
\end{eqnarray}
Compared to Eq. (17) this only suggests an improved correction by a factor $1/2$. 

\section{Regularised self-energy due to the electric field}

In T-duality, it was shown that for the electric potential of the system which consists of two point-like masses, at relative distance $\vec{r}$, one can write  \cite{Gaete:2022une}
\begin{align}
V_{em}(r) \to -\frac{Q}{4\pi \epsilon_0 \sqrt{r^2 + l_0^2}}.
\end{align}

 In addition, we have Maxwell's equation which is written as 
 \begin{equation}
 \nabla_{\mu}F^{\mu \nu}= J^{\nu},
 \end{equation}
 where $F_{\mu\nu}=\nabla_\mu A_{\nu}-\nabla_\nu A_{\mu}$ and $J^{\mu}=(c \rho_{em},\vec{j})$ is the four-current. One can check that the above equations lead to the charge density
 \begin{equation}
     \rho_{em}(r)= \frac{3 l_0^2 Q}{ { (r^2 +l_0^2)}^{5/2}}.
 \end{equation} 
 The interaction Hamiltonian then becomes
\begin{eqnarray}
\label{eqn:hint}
 H^{em}_\text{int} &=& -Q\int_{0}^{2 \pi}\int_{0}^{\pi}\int_0^{\infty} \mathrm{d}^3 r \frac{\rho_{em}(\vec r)}{4 \pi \epsilon_0 \sqrt{r^2+l_0^2}} \,,
\end{eqnarray}
Solving the integral we get
\begin{equation}
     H^{em}_\text{int}=-\frac{3 \pi Q^2}{16 \epsilon_0 l_0}.
\end{equation}
\subsection{Curved spacetime}
If we consider again the effect of spacetime curvature, by using the mass profile with corrections due to the charge density (36) we obtain
\begin{equation}
f(r)= 1 - \frac{2 Q r^2}{{\left(r^2 +l_0^2\right)}^{3/2}}=\left(1+2\Phi_{em}(r)\right),
\end{equation}
which does not represent a new solution since the total mass is simply scaled to $M\to M + Q$. The Schr\"odinger--Newton equation will be
\begin{equation}
i \hbar \frac{\partial \Psi(t,\vec r) }{\partial t}= \left( -\frac{\hbar^2}{2 M} \nabla_{LB}^2+H_{int}+H_{int}^{em} \right)\Psi(t,\vec r).
\end{equation}
The interaction Hamiltonian due to the charge density we get
\begin{eqnarray}
\label{eqn:hint}
 H^{em}_\text{int} &=& \int_{0}^{2 \pi}\int_{0}^{\pi}\int_0^{\infty} \mathrm{d}^3 r\, \Phi_{em}(r){\rho}_{em},
\end{eqnarray}
yielding
\begin{equation}
     H^{em}_\text{int}=-\frac{3 \pi Q^2}{32 \epsilon_0 l_0}
\end{equation}
Compared to (38), we obtain an improved correction by a factor of $1/2$. The total energy of the particle is
\begin{equation}
    E_{tot}=E+\frac{3 \pi G M^2}{32 l_0}+\frac{3 \pi Q^2}{32  \epsilon_0 l_0}.
\end{equation}
This equation shows that the electrostatic field also modifies the energy of the particle. Here we found the energy stored in the charge via the charged density and the curved spacetime. It is very interesting to note that the corrections to the energy due to the electromagnetic field can be found in an alternative way using the stress-energy tensor for the electromagnetic field. We will assume that the form of the energy-tensor the electromagnetic field has the form 
\begin{equation}
 T_{\mu \nu}^{em}=F_{\mu \sigma}{F_{\nu}}^{\sigma}    -\frac{1}{4}g_{\mu \nu}F_{\rho \sigma}F^{\rho \sigma},
\end{equation}
where 
\begin{eqnarray}
    A_{\mu}=\left(\frac{V_{em}}{c},0,0,0\right).
\end{eqnarray}
It is easy to show that the ${T^0}_{0}$ component of the energy-momentum tensor for the electromagnetic field gives the energy density  
\begin{equation}
    \rho_{em}=\frac{Q^2 r^2}{8 \pi (r^2+l_0^2)^3}.
\end{equation}
This means that the energy has to be 
\begin{eqnarray}
    H_{int}^{em}&=-&\int \mathrm{d}^3r \frac{Q^2 r^2}{8 \pi (r^2+l_0^2)^3}\\
    &=-&\frac{3\pi  Q^2}{32 \epsilon_0 l_0 },
\end{eqnarray}
 which precisely matches Eq. (42). Such a correction was found also for a charged black hole in T-duality \cite{Gaete:2022ukm}.

 \section{Modified Klein-Gordon equation}
 We can obtain the corrected relativistic energy for the particle with a different spin. Let us consider here the Klein-Gordon described by the scalar field $\Phi(x,t)$
 \begin{eqnarray}
     \square\Phi(x,t)-\frac{M^2 c^2}{\hbar^2}\Phi(x,t)-\frac{H^2_{int}}{\hbar^2 c^2}\Phi(x,t)=0,
 \end{eqnarray}
 which is modified due to the self-gravitational interaction term. 
This equation can be further written as
 \begin{eqnarray}
     \left(\nabla^2-\frac{1}{c^2}\frac{\partial^2}{\partial t^2}-\frac{M^2 c^2}{\hbar^2}\right)\Phi(x,t)-\frac{H^2_{int}}{\hbar^2 c^2}\Phi(x,t)=0.
 \end{eqnarray}
 Taking a plane wave solution for the scalar field
\begin{equation}
    \Phi(x,t)=A \exp(-\frac{i}{\hbar}x^{\mu}p_{\mu}),
\end{equation}
 we get the energy relation given by 
 \begin{eqnarray}
    E^2_{tot}=M^2c^4+p^2 c^2+\frac{9 \pi^2 G^2 M^4}{16^2 l_0^2 }.
\end{eqnarray}
We can also rewrite this equation in a more compact form
\begin{eqnarray}
    E^2_{tot}=M_{tot}^2c^4+p^2 c^2.
\end{eqnarray}
in terms of the new definition of the mass
\begin{eqnarray}
    M^2_{tot}=M^2\left(1+\frac{9 \pi^2 G^2 M^2 }{16^2 c^4 l_0^2}\right).
\end{eqnarray}
By considering a series expansion, we get
\begin{eqnarray}
    M_{tot}\simeq M\left(1+\frac{1}{2}\frac{9 \pi^2 G^2 M^2}{16^2 c^4 l_0^2 }\right).
\end{eqnarray}
Using the fact that $l_0 \sim l_{Pl}=\sqrt{\frac{\hbar G}{c^3}}$, we can also write 
\begin{eqnarray}
    M_{tot}\simeq M\left(1+\alpha \frac{ M^2}{M_{Pl}^2}\right).
\end{eqnarray}
where $\alpha=(9/2)\,\pi^2/16^2$ and $M_{Pl}=\sqrt{\frac{\hbar c}{G}}$ is the Planck mass. This expression matches the one derived in the extended uncertainty principle derived in  \cite{Mureika:2018gxl}. This could be related to gravitational self-completeness with quantum mechanical mass limits. In  particular, we may have two cases: \\
1) When $M<<M_{Pl}$, i.e., in the particle sector then Eq. (56) is valid and the corrections are small. \\
2) When $M>M_{Pl}$, we have a connection between elementary
particles and black holes. As was shown \cite{Carr:2015nqa,Carr:2022ndy} the Compton-Schwarzschild correspondence posits a smooth transition between the Compton wavelength $(R_C \sim 1/M)$ below the Planck mass and the Schwarzschild radius $(R_S \sim M)$ above it. The Compton and Schwarzschild lines transform into one another under the
transformation $M \to M^2_{Pl}/M$, which suggests the following form for the mass 
\begin{eqnarray}
    M_{tot}\simeq M\left(1+\alpha \frac{M_{Pl}^2}{M^2}\right).
\end{eqnarray}
This relation was recently obtained in Refs. \cite{Carr:2015nqa,Carr:2022ndy}.  

\section{Modified Dirac equation}
Let us consider also here the Dirac equation described by the spinor field $\psi(x,t)$ which is modified due to the self-gravitational interaction term
 \begin{eqnarray}
    \mathrm{i} \gamma^{\mu} \nabla_{\mu} \psi(x,t)-\frac{M c}{\hbar}\psi(x,t)+\frac{H_{int}}{\hbar c}\psi(x,t)=0,
 \end{eqnarray}
where for the $\gamma^\mu$ matrices in terms of $2 \times 2$ Pauli sub-matrices we can write 
\begin{eqnarray}
    \gamma^{t}=\left( \begin{matrix}
 I_2 &  0 \\
0 & -I_2 
\end{matrix}\right),\,\,\, \gamma^{1}=\left( \begin{matrix}
 0 &  \sigma_x \\
-\sigma_x & 0 
\end{matrix}\right),
\end{eqnarray}
\begin{eqnarray}
   \gamma^{2}=\left( \begin{matrix}
 0 &  \sigma_y \\
-\sigma_y & 0 
\end{matrix}\right),\,\,\, \gamma^{3}=\left( \begin{matrix}
 0 &  \sigma_z \\
-\sigma_z & 0 
\end{matrix}\right).
\end{eqnarray}

If we take the solution for the spinor field 
\begin{eqnarray}
    \psi(x,t)=\left( \begin{matrix}
\psi_1 \\
 \psi_2\\
 \psi_3\\
 \psi_4
\end{matrix} \right),
\end{eqnarray}
or, in a different notation 
\begin{eqnarray}
    \psi(x,t)=\left( \begin{matrix}
\psi_A \\
 \psi_B\\
\end{matrix} \right),\,\,\,  \psi_A=\left( \begin{matrix}
\psi_1 \\
 \psi_2\\
\end{matrix} \right),\,\,\,\psi_B=\left( \begin{matrix}
\psi_3 \\
 \psi_4\\
\end{matrix} \right).
\end{eqnarray}
It is well known that the solution can be written as 
\begin{eqnarray}
    \psi(x^{\mu})=A \exp\left(-\frac{\mathrm{i}}{\hbar} x^{\mu} p_{\mu}\right)\left( \begin{matrix}
u_A \\
 u_B\\
\end{matrix} \right)
\end{eqnarray}
then the Dirac equation reduces to 
\begin{eqnarray}
    \left(\gamma^{\mu} p_{\mu}-Mc +\frac{H_{int}}{c}\right)\left( \begin{matrix}
u_A \\
 u_B\\
\end{matrix} \right)=0,
\end{eqnarray}
From this equation, we get a system of two equations 
\begin{eqnarray}
    \left( \begin{matrix}
 \frac{E}{c}-Mc+\frac{H_{int}}{c} &  - \vec p \cdot \vec\sigma\\
\vec p \cdot \vec\sigma & -\frac{E}{c}-Mc+\frac{H_{int}}{c}
\end{matrix}\right)\left( \begin{matrix}
u_A \\
 u_B\\
\end{matrix} \right)=0.
\end{eqnarray}
By solving the determinant of this system we get two solutions for the energy 
\begin{eqnarray}
    E_{tot}=\pm \sqrt{M_{tot}^2 c^4+p^2 c^2},
\end{eqnarray}
where it has been defined
\begin{eqnarray}
    M^2_{tot}c^4=\left(M c^2+\frac{3 \pi G M^2}{16 l_0 }\right)^2.
\end{eqnarray}

Finally, the last equation can be further written as
\begin{eqnarray}
    M_{tot}=M\left(1+\frac{3 \pi G M}{16 c^2 l_0 }\right),
\end{eqnarray}
or 
\begin{eqnarray}
    M_{tot}=M\left(1+\alpha \frac{ M}{M_{Pl}}\right),
\end{eqnarray}
where $\alpha=3 \pi/16$. It is rather amazing that the correction to the mass for fermions (spin 1/2) is linear as can be seen from the last equation. On the other hand bosons (spin zero) is quadratic as was found in Eq. (56). This is a surprising result. Again we have two cases:\\
1) When $M<<M_{Pl}$, then Eq. (69) is valid for the particle sector. \\
2) When $M>M_{Pl}$, we have a connection between elementary
particles and black holes and we have to use the 
transformation $M \to M^2_{Pl}/M$ \cite{Carr:2015nqa,Carr:2022ndy}, which suggests the following form for the mass 
\begin{eqnarray}
    M_{tot}\simeq M\left(1+\alpha \frac{M_{Pl}}{M}\right).
\end{eqnarray}

\section{Modified Proca equation}
Here we shall consider the motion of a massive vector particle of mass $M$,  with spin $1$ described by the vector field
$\psi^{\mu}$, which can be studied by the Proca equation (PE), which reads 
\be\label{Proca}
\nabla_{\mu}\nabla^{[\mu}\psi^{\nu]}-\frac{M^{2}c^2}{\hbar^{2}}\psi^{\nu}-\frac{H_{int}^{2}}{\hbar^{2} c^2}\psi^{\nu}=0
\ee
where in the last term we introduced the corrections due to the self-regularized energy. Note also that 
\be
\nabla_{[\mu}\psi_{\nu]}=\frac{1}{2}(\nabla_{\mu}\psi_{\nu}-\nabla_{\nu}\psi_{\mu}):=\psi_{\mu\nu}~.
\ee

The solution of the equation can be written as
\be
\psi_{\nu}(x^{\mu})=C_{\nu}\exp\left(-\frac{\mathrm{i}}{\hbar} x^{\mu} p_{\mu}\right)
\ee
 Using  the four-momentum $p^{\mu}=(E/c, \vec p)$  we can construct a $4\times4$ matrix $\aleph$, which satisfies the following matrix equation 
\be
 \aleph(C_{0},C_{1},C_{2},C_{3})^{T}=0~.
\ee

Taking for simplicity only one component of the momentum $\vec p=(p_x,0,0)$, the matrix $\aleph$ has the components 
\begin{eqnarray}
    \aleph=\left( \begin{matrix}
   \aleph_{00} &  0  & 0 &   \aleph_{03}\\
  \aleph_{10} &  0  & 0 &   \aleph_{13}\\
0 &     \aleph_{21}  & 0 & 0 \\
0 &  0  &    \aleph_{32} & 0 \\
\end{matrix}\right)
\end{eqnarray}
where 
\begin{eqnarray}
  \aleph_{00}&=& \frac{E p}{c^2 \hbar^2},\\
  \aleph_{03}&=&-\frac{M^2 c^4+c^2 p^2-H_{int}^2}{c^2},\\
   \aleph_{10}&=& \frac{M^2 c^4-E^2+H_{int}^2}{c^2},\\
   \aleph_{13}&=&-\frac{E p}{c^2 },\\
   \aleph_{21}&=&\aleph_{32}=\frac{M^2 c^4+c^2 p^2-E^2+H_{int}^2}{c^2}. 
\end{eqnarray}
Taking the determinant of this matrix and solving for the energy we get
\begin{eqnarray}
    E_{tot}=\pm \sqrt{M_{tot}^2 c^4+p^2 c^2}
\end{eqnarray}
where 
\begin{eqnarray}
    M^2_{tot}=M^2\left(1+\frac{9 \pi^2 G^2 M^2 }{16^2 c^4 l_0^2}\right),
\end{eqnarray}
and therefore
\begin{eqnarray}
    M_{tot}\simeq M\left(1+\alpha \frac{ M^2}{M_{Pl}^2}\right).
\end{eqnarray}
In other words, the corrections to the energy and mass of the particle with spin 1 particles described by the Proca equation are similar to scalar particles with spin 0 described by the Klein-Gordon equation. 

\section{GUP from regularized self-energy}
According to the non-commutative relation between position and momentum, we have
\begin{eqnarray}
\left[x_i,p_j\right]= \mathrm{i}~\delta_{ij}~\hbar,
\end{eqnarray}
meaning that the position and momentum can not have real eigenvalue for the same eigenstate. In quantum gravity theories, such as string theory \cite{Amati:1988tn} the quadratic form of GUP has been suggested and has the following form \cite{Kempf:1994su,Brau:1999uv,Maggiore:1993rv}:
\begin{eqnarray}
\label{gupquadratic}
\Delta x\Delta p &\geq& \frac{\hbar}{2}(1+\beta\Delta p^2), \nonumber\\
\left[x_i,p_j\right]&=&\mathrm{i}\hbar\left[\delta_{ij}+\beta \delta_{ij}p^2+2\beta p_i p_j\right]
\end{eqnarray}
where $\beta=\beta_0 l_{Pl}^2/\hbar^2$, $\beta_0$ is a dimensionless constant, and $l_{Pl}=1.6162\times 10^{-35}\text{m}$ is the Planck length. Besides, the form of linear GUP  is also motivated by doubly special relativity \cite{Magueijo:2001cr,Magueijo:2002am}. The linear GUP has the following form\cite{Ali:2009zq,Cortes:2004qn,Ali:2011fa}
\begin{eqnarray}
[x_i, p_j] = \mathrm{i} \hbar\hspace{-0.5ex} \left[  \delta_{ij}\hspace{-0.5ex}
- \hspace{-0.5ex} \alpha\hspace{-0.5ex}  \left( p \delta_{ij} +
\frac{p_i p_j}{p} \right)
 \right],~~~
\end{eqnarray}
where $\alpha=\alpha_0 l_{Pl}/\hbar$, and $\alpha_0$ is a dimensionless constant.  In what follows we shall argue how these two types of GUP  appear naturally from regularized self-energy.

\subsection{Quadratic GUP and bosons particles}
Let us show a very interesting result where one can obtain the GUP principle using the modified energy.  Let us rewrite Eqs. (56) and (83) as follows 
\begin{eqnarray}
     E_{tot}\simeq E\left(1+\alpha \frac{ E^2}{c^4 M_{Pl}^2}\right).
\end{eqnarray}
where $E_{tot}=M_{tot} c^2$ and $E=M c^2$. By differentiating  with respect to the energy on both sides we get 
\begin{eqnarray}
    \mathrm{d}E_{tot} \simeq \mathrm{d}E\left(1+3 \alpha \frac{ E^2}{c^4 M_{Pl}^2}\right).
\end{eqnarray}
Only in the limit $\alpha \to 0$, for energies we have $E_{tot}=E$, therefore, in general, we expect two possibilities. The first possibility is to assume 
\begin{eqnarray}
    \mathrm{d}E_{tot}\mathrm{d}t \sim \hbar.
\end{eqnarray}
If we multiply Eq. (87) in both sides with $dt$ we then get in leading order terms
\begin{eqnarray}
    \mathrm{d}E \mathrm{d}t \sim \hbar \left(1- 3 \alpha \frac{ E^2}{c^4 M_{Pl}^2}\right),
\end{eqnarray}
but since 
\begin{eqnarray}
    l_0 \sim l_{Pl}=\sqrt{\frac{\hbar G}{c^3}},\,\,\,   M_{Pl}=\sqrt{\frac{\hbar c}{G}},
\end{eqnarray}
we can write Eq. (90) further as
\begin{eqnarray}
    \mathrm{d}E \mathrm{d}t \sim \hbar \left(1-\frac{\beta\,  l^2_{Pl} E^2}{ c^2 \hbar^2}\right),
\end{eqnarray}
where $\beta=3\alpha$. The second possibility is to assume
    \begin{eqnarray}
    \mathrm{d}E \mathrm{d}t \sim \hbar,
\end{eqnarray}
and, again, by multiplying Eq. (87) with $dt$ we get
\begin{eqnarray}
    \mathrm{d}E_{tot} \mathrm{d}t \sim \hbar \left(1+\frac{\beta\,  l^2_{Pl} E^2}{ c^2 \hbar^2}\right).
\end{eqnarray}
In other words, there are two ways of expressing the GUP corrected time-energy relation with the difference in the sign. The difference in the sign before the second term is of course related to whether we choose to work with the total or the bare energy.  

This shows that the GUP for the time-energy can be viewed as a consequence of the modified energy of the particle due to the regularized self-energy. To obtain the momentum-time GUP relation, we simply need to use Eqs. (56) and (83) and making use of $\Delta p_{tot} \to c M_{tot}$, along with $\Delta p \to c M$ (similar relations have been used for example in \cite{Carr:2015nqa}), and we multiply both sides by $\Delta x$, we get 
\begin{eqnarray}
    \Delta p_{tot}\,\Delta x \sim \Delta p\, \Delta x \left(1+\frac{\alpha\,  l^2_{Pl}\, \Delta p^2}{ \hbar^2}\right).
\end{eqnarray}
We have a similar situation where $\Delta p_{tot}=\Delta p$, provided $\alpha \to 0$. To this end, we can first assume $ \Delta p_{tot}\,\Delta x \sim \hbar$. In that case we obtain
\begin{eqnarray}
    \Delta p\,\Delta x \sim \hbar\, \left(1-\frac{\alpha\,  l^2_{Pl}\, \Delta p^2}{ \hbar^2}\right).
\end{eqnarray}
This GUP expression is consistent with the quadratic form of GUP given by Eq. (85), provided $\alpha=-\beta_0/2$. The second case is to assume $ \Delta p\,\Delta x \sim \hbar$, then from Eq. (95) we obtain
\begin{eqnarray}
    \Delta p_{tot}\,\Delta x \sim \hbar\, \left(1+\frac{\alpha\,  l^2_{Pl}\, \Delta p^2}{ \hbar^2}\right).
\end{eqnarray}
We, therefore, found that in general there are two equivalent representations of the GUP corrected momentum-position relation. In general $\beta_0$ is a free parameter that can be positive or negative. This sign in the second term is also related to the fact whether we work with $\Delta p_{tot}$ or $\Delta p$. 

\subsection{Linear GUP and Dirac particles}
Consider the corrections to the mass given by Eq. (69) which can be written as
\begin{eqnarray}
     E_{tot}\simeq E\left(1+\alpha \frac{ E}{c^2 M_{Pl}}\right).
\end{eqnarray}
By differentiating  with respect to the energy on both sides we get 
\begin{eqnarray}
    \mathrm{d}E_{tot}=\mathrm{d}E\left(1+ 2\alpha \frac{ E}{c^2 M_{Pl}}\right).
\end{eqnarray}
As in the above discussion, let us consider first the case 
\begin{eqnarray}
    \mathrm{d}E_{tot}\mathrm{d}t \sim \hbar,
\end{eqnarray}
we get 
\begin{eqnarray}
    \mathrm{d}E \mathrm{d}t \sim \hbar \left(1-2\alpha \frac{ E}{c^2 M_{Pl}}\right),
\end{eqnarray}
we can write further this as 
\begin{eqnarray}
    \mathrm{d}E \mathrm{d}t \sim \hbar \left(1-\frac{\beta\,  l_{Pl} E}{ c \hbar}\right),
\end{eqnarray}
where $\beta=2\alpha$. To get the second representation we can assume 
\begin{eqnarray}
    \mathrm{d}E \mathrm{d}t \sim \hbar,
\end{eqnarray}
in Eq. (98). If follows then 
\begin{eqnarray}
    \mathrm{d}E_{tot} \mathrm{d}t \sim \hbar \left(1+\frac{\beta\,  l_{Pl} E}{ c \hbar}\right).
\end{eqnarray}
This shows again that in both cases the time-energy GUP relation can be a consequence of the modified energy of the particle due to the regularized self-energy. If we further take  $\Delta p_{tot} \to c M_{tot}$ and $\Delta p \to c M$  in Eq. (69), we obtain 
\begin{eqnarray}
    \Delta p_{tot}\,\Delta x \sim \Delta p\, \Delta x \left(1+\frac{\alpha\,  l_{Pl}\, \Delta p}{ \hbar}\right),
\end{eqnarray}
again by means of $ \Delta p_{tot}\,\Delta x \sim \hbar$, we obtain
\begin{eqnarray}
    \Delta p\,\Delta x \sim \hbar\, \left(1-\frac{\alpha\,  l_{Pl}\, \Delta p}{ \hbar}\right).
\end{eqnarray}
Finally,  if we assume $ \Delta p\,\Delta x \sim \hbar$ in Eq. (105), we the modified GUP relation
\begin{eqnarray}
    \Delta p_{tot}\,\Delta x \sim \hbar\, \left(1+\frac{\alpha\,  l_{Pl}\, \Delta p}{ \hbar}\right).
\end{eqnarray}

This is the linear GUP in agreement with Eq. (86). We should point out that by comparing it to Eq. (86), the correction can be positive or negative. This has to do with the fact that the free parameter $\alpha_0$ can be positive or negative.  We, therefore, find that the form of GUP depends on the nature of the particles being considered. In particular, for scalar particles with spin zero, we obtained a quadratic form of GUP, while for fermion particles with spin 1/2, we got a linear GUP. This is consistent with the connection between spin and linear GUP that was found in \cite{Ali:2021oml} that have implications with explaining the quantum entanglement \cite{ Ali:2022jna} through finding a relation between uncertainty and Bekenstein bound \cite{Buoninfante:2020guu}.

\section{Conclusions}
In the present paper, we have computed the regularized self-energy of the particle using the Schr\"odinger--Newton equation. We used regular and well-defined gravitational and electric potentials obtained in T-duality. In this picture, the particle with mass $M$ is no longer concentrated into a point but it is diluted and can be described by a quantum-corrected stress-energy tensor with a smeared energy density resulting in corrections to the energy of the particle which can be interpreted as a regularized self-energy of the particle. To this end, we have found the corrections to the energy due to the electrostatic field or the energy stored in the charge density of the particle. 

In the second part of this work, we have extended the corrections to energy by incorporating the relativistic effects using the relativistic field equations such as the Klein-Gordon, Proca, and Dirac equations.  We found that the corrections to the energy can be linked to the GUP. Quite surprisingly, the form of GUP is shown to depend on the spin of the particles; namely, for bosons having spin $0$ and $1$ we obtain a quadratic form of GUP, on the other hand for fermions having spin $1/2$ we obtain a linear form of GUP. In the near future, we would like to study more about the phenomenological aspects of GUP which is linked to regularized self-energy as was argued in this work. We would like to see how the regularized self-energy affects other types of particles such as the spin-2 particles. Like spin-0 and spin-1 particles, one can of course assume also for gravitons a similar  expression for the energy correction or modified dispersion relation 
\begin{eqnarray}
    E_{tot}^2=(p^g)^2 c^2+(M^g)^2_{tot} c^4,
\end{eqnarray}
where $M^g_{tot}\simeq M^g \left(1+\alpha\, (M^g/M_{Pl})^2\right)$, with $M^g$ and $p^g$ being the bare graviton mass and the graviton momentum, respectively. Using the expression for the gravitons three-velocity $v^g=p^g\,c^2/E_{tot}$, we can obtain 
\begin{eqnarray}
    \frac{v^g}{c}=\sqrt{1-\frac{(M^g_{tot})^2c^4}{E_{tot}^2}}.
\end{eqnarray}
For the energy we may take $E_{tot}=h f$, with $f$ being the
graviton’s frequency, then using the gravitational waves it might be interesting to see whether one can constrain the graviton mass along with the zero-point length. There is also this possibility that we cannot constrain the zero-point length from observations. This has to do with the fact that $M_{tot}^g$ might be the true mass being measured by observations and, it simply shifts by some constant compared to the bare mass $M^g$, meaning that we cannot distinguish these quantities with observations.  We would like to study more about the implication of GUP on the graviton mass in the near future. 

\bibliography{ref}

\begin{thebibliography}{67}%
\makeatletter
\providecommand \@ifxundefined [1]{%
 \@ifx{#1\undefined}
}%
\providecommand \@ifnum [1]{%
 \ifnum #1\expandafter \@firstoftwo
 \else \expandafter \@secondoftwo
 \fi
}%
\providecommand \@ifx [1]{%
 \ifx #1\expandafter \@firstoftwo
 \else \expandafter \@secondoftwo
 \fi
}%
\providecommand \natexlab [1]{#1}%
\providecommand \enquote  [1]{``#1''}%
\providecommand \bibnamefont  [1]{#1}%
\providecommand \bibfnamefont [1]{#1}%
\providecommand \citenamefont [1]{#1}%
\providecommand \href@noop [0]{\@secondoftwo}%
\providecommand \href [0]{\begingroup \@sanitize@url \@href}%
\providecommand \@href[1]{\@@startlink{#1}\@@href}%
\providecommand \@@href[1]{\endgroup#1\@@endlink}%
\providecommand \@sanitize@url [0]{\catcode `\\12\catcode `\$12\catcode
  `\&12\catcode `\#12\catcode `\^12\catcode `\_12\catcode `\%12\relax}%
\providecommand \@@startlink[1]{}%
\providecommand \@@endlink[0]{}%
\providecommand \url  [0]{\begingroup\@sanitize@url \@url }%
\providecommand \@url [1]{\endgroup\@href {#1}{\urlprefix }}%
\providecommand \urlprefix  [0]{URL }%
\providecommand \Eprint [0]{\href }%
\providecommand \doibase [0]{http://dx.doi.org/}%
\providecommand \selectlanguage [0]{\@gobble}%
\providecommand \bibinfo  [0]{\@secondoftwo}%
\providecommand \bibfield  [0]{\@secondoftwo}%
\providecommand \translation [1]{[#1]}%
\providecommand \BibitemOpen [0]{}%
\providecommand \bibitemStop [0]{}%
\providecommand \bibitemNoStop [0]{.\EOS\space}%
\providecommand \EOS [0]{\spacefactor3000\relax}%
\providecommand \BibitemShut  [1]{\csname bibitem#1\endcsname}%
\let\auto@bib@innerbib\@empty
\bibitem [{\citenamefont {Einstein}(1916)}]{Einstein:1916vd}%
  \BibitemOpen
  \bibfield  {author} {\bibinfo {author} {\bibfnamefont {Albert}\ \bibnamefont
  {Einstein}},\ }\bibfield  {title} {\enquote {\bibinfo {title} {{The
  Foundation of the General Theory of Relativity}},}\ }\href {\doibase
  10.1002/andp.19163540702} {\bibfield  {journal} {\bibinfo  {journal} {Annalen
  Phys.}\ }\textbf {\bibinfo {volume} {49}},\ \bibinfo {pages} {769--822}
  (\bibinfo {year} {1916})}\BibitemShut {NoStop}%
\bibitem [{\citenamefont {Schr\"odinger}(1926)}]{Schrodinger:1926gei}%
  \BibitemOpen
  \bibfield  {author} {\bibinfo {author} {\bibfnamefont {E.}~\bibnamefont
  {Schr\"odinger}},\ }\bibfield  {title} {\enquote {\bibinfo {title}
  {{Quantisierung als Eigenwertproblem}},}\ }\href {\doibase
  10.1002/andp.19263840404} {\bibfield  {journal} {\bibinfo  {journal} {Annalen
  Phys.}\ }\textbf {\bibinfo {volume} {384}},\ \bibinfo {pages} {361--376}
  (\bibinfo {year} {1926})}\BibitemShut {NoStop}%
\bibitem [{\citenamefont {Gordon}(1926)}]{gordon1926comptoneffekt}%
  \BibitemOpen
  \bibfield  {author} {\bibinfo {author} {\bibfnamefont {Walter}\ \bibnamefont
  {Gordon}},\ }\bibfield  {title} {\enquote {\bibinfo {title} {Der
  comptoneffekt nach der schr{\"o}dingerschen theorie},}\ }\href@noop {}
  {\bibfield  {journal} {\bibinfo  {journal} {Zeitschrift f{\"u}r Physik}\
  }\textbf {\bibinfo {volume} {40}},\ \bibinfo {pages} {117--133} (\bibinfo
  {year} {1926})}\BibitemShut {NoStop}%
\bibitem [{\citenamefont {Klein}(1926)}]{klein1926quantentheorie}%
  \BibitemOpen
  \bibfield  {author} {\bibinfo {author} {\bibfnamefont {Oskar}\ \bibnamefont
  {Klein}},\ }\bibfield  {title} {\enquote {\bibinfo {title} {Quantentheorie
  und f{\"u}nfdimensionale relativit{\"a}tstheorie},}\ }\href@noop {}
  {\bibfield  {journal} {\bibinfo  {journal} {Zeitschrift f{\"u}r Physik}\
  }\textbf {\bibinfo {volume} {37}},\ \bibinfo {pages} {895--906} (\bibinfo
  {year} {1926})}\BibitemShut {NoStop}%
\bibitem [{\citenamefont {Dirac}(1928)}]{Dirac:1928hu}%
  \BibitemOpen
  \bibfield  {author} {\bibinfo {author} {\bibfnamefont {Paul A.~M.}\
  \bibnamefont {Dirac}},\ }\bibfield  {title} {\enquote {\bibinfo {title} {{The
  quantum theory of the electron}},}\ }\href {\doibase 10.1098/rspa.1928.0023}
  {\bibfield  {journal} {\bibinfo  {journal} {Proc. Roy. Soc. Lond. A}\
  }\textbf {\bibinfo {volume} {117}},\ \bibinfo {pages} {610--624} (\bibinfo
  {year} {1928})}\BibitemShut {NoStop}%
\bibitem [{\citenamefont {Di\'osi}(1984)}]{Diosi:1984wuz}%
  \BibitemOpen
  \bibfield  {author} {\bibinfo {author} {\bibfnamefont {L.}~\bibnamefont
  {Di\'osi}},\ }\bibfield  {title} {\enquote {\bibinfo {title} {{Gravitation
  and quantummechanical localization of macroobjects}},}\ }\href {\doibase
  10.1016/0375-9601(84)90397-9} {\bibfield  {journal} {\bibinfo  {journal}
  {Phys. Lett. A}\ }\textbf {\bibinfo {volume} {105}},\ \bibinfo {pages}
  {199--202} (\bibinfo {year} {1984})},\ \Eprint
  {http://arxiv.org/abs/1412.0201} {arXiv:1412.0201 [quant-ph]} \BibitemShut
  {NoStop}%
\bibitem [{\citenamefont {Diosi}(1987)}]{Diosi:1986nu}%
  \BibitemOpen
  \bibfield  {author} {\bibinfo {author} {\bibfnamefont {L.}~\bibnamefont
  {Diosi}},\ }\bibfield  {title} {\enquote {\bibinfo {title} {{A Universal
  Master Equation for the Gravitational Violation of Quantum Mechanics}},}\
  }\href {\doibase 10.1016/0375-9601(87)90681-5} {\bibfield  {journal}
  {\bibinfo  {journal} {Phys. Lett. A}\ }\textbf {\bibinfo {volume} {120}},\
  \bibinfo {pages} {377} (\bibinfo {year} {1987})}\BibitemShut {NoStop}%
\bibitem [{\citenamefont {Penrose}(1996)}]{Penrose:1996cv}%
  \BibitemOpen
  \bibfield  {author} {\bibinfo {author} {\bibfnamefont {Roger}\ \bibnamefont
  {Penrose}},\ }\bibfield  {title} {\enquote {\bibinfo {title} {{On gravity's
  role in quantum state reduction}},}\ }\href {\doibase 10.1007/BF02105068}
  {\bibfield  {journal} {\bibinfo  {journal} {Gen. Rel. Grav.}\ }\textbf
  {\bibinfo {volume} {28}},\ \bibinfo {pages} {581--600} (\bibinfo {year}
  {1996})}\BibitemShut {NoStop}%
\bibitem [{\citenamefont {Diosi}(1989)}]{Diosi:1988uy}%
  \BibitemOpen
  \bibfield  {author} {\bibinfo {author} {\bibfnamefont {L.}~\bibnamefont
  {Diosi}},\ }\bibfield  {title} {\enquote {\bibinfo {title} {{MODELS FOR
  UNIVERSAL REDUCTION OF MACROSCOPIC QUANTUM FLUCTUATIONS}},}\ }\href@noop {}
  {\bibfield  {journal} {\bibinfo  {journal} {Phys. Rev. A}\ }\textbf {\bibinfo
  {volume} {40}},\ \bibinfo {pages} {1165--1174} (\bibinfo {year}
  {1989})}\BibitemShut {NoStop}%
\bibitem [{\citenamefont {Penrose}(1998)}]{penrose1998quantum}%
  \BibitemOpen
  \bibfield  {author} {\bibinfo {author} {\bibfnamefont {Roger}\ \bibnamefont
  {Penrose}},\ }\bibfield  {title} {\enquote {\bibinfo {title} {Quantum
  computation, entanglement and state reduction},}\ }\href@noop {} {\bibfield
  {journal} {\bibinfo  {journal} {Philosophical Transactions of the Royal
  Society of London. Series A: Mathematical, Physical and Engineering
  Sciences}\ }\textbf {\bibinfo {volume} {356}},\ \bibinfo {pages} {1927--1939}
  (\bibinfo {year} {1998})}\BibitemShut {NoStop}%
\bibitem [{\citenamefont {Penrose}(2014)}]{Penrose:2014nha}%
  \BibitemOpen
  \bibfield  {author} {\bibinfo {author} {\bibfnamefont {Roger}\ \bibnamefont
  {Penrose}},\ }\bibfield  {title} {\enquote {\bibinfo {title} {{On the
  Gravitization of Quantum Mechanics 1: Quantum State Reduction}},}\ }\href
  {\doibase 10.1007/s10701-013-9770-0} {\bibfield  {journal} {\bibinfo
  {journal} {Found. Phys.}\ }\textbf {\bibinfo {volume} {44}},\ \bibinfo
  {pages} {557--575} (\bibinfo {year} {2014})}\BibitemShut {NoStop}%
\bibitem [{\citenamefont {Proca}(1936)}]{Proca:1936fbw}%
  \BibitemOpen
  \bibfield  {author} {\bibinfo {author} {\bibfnamefont {Alexandru}\
  \bibnamefont {Proca}},\ }\bibfield  {title} {\enquote {\bibinfo {title} {{Sur
  la theorie ondulatoire des electrons positifs et negatifs}},}\ }\href
  {\doibase 10.1051/jphysrad:0193600708034700} {\bibfield  {journal} {\bibinfo
  {journal} {J. Phys. Radium}\ }\textbf {\bibinfo {volume} {7}},\ \bibinfo
  {pages} {347--353} (\bibinfo {year} {1936})}\BibitemShut {NoStop}%
\bibitem [{\citenamefont {Sathiapalan}(1987)}]{Sathiapalan:1986zb}%
  \BibitemOpen
  \bibfield  {author} {\bibinfo {author} {\bibfnamefont {B.}~\bibnamefont
  {Sathiapalan}},\ }\bibfield  {title} {\enquote {\bibinfo {title} {{Duality in
  Statistical Mechanics and String Theory}},}\ }\href {\doibase
  10.1103/PhysRevLett.58.1597} {\bibfield  {journal} {\bibinfo  {journal}
  {Phys. Rev. Lett.}\ }\textbf {\bibinfo {volume} {58}},\ \bibinfo {pages}
  {1597} (\bibinfo {year} {1987})}\BibitemShut {NoStop}%
\bibitem [{\citenamefont {Strominger}\ \emph {et~al.}(1996)\citenamefont
  {Strominger}, \citenamefont {Yau},\ and\ \citenamefont
  {Zaslow}}]{Strominger:1996it}%
  \BibitemOpen
  \bibfield  {author} {\bibinfo {author} {\bibfnamefont {Andrew}\ \bibnamefont
  {Strominger}}, \bibinfo {author} {\bibfnamefont {Shing-Tung}\ \bibnamefont
  {Yau}}, \ and\ \bibinfo {author} {\bibfnamefont {Eric}\ \bibnamefont
  {Zaslow}},\ }\bibfield  {title} {\enquote {\bibinfo {title} {{Mirror symmetry
  is T duality}},}\ }\href {\doibase 10.1016/0550-3213(96)00434-8} {\bibfield
  {journal} {\bibinfo  {journal} {Nucl. Phys. B}\ }\textbf {\bibinfo {volume}
  {479}},\ \bibinfo {pages} {243--259} (\bibinfo {year} {1996})},\ \Eprint
  {http://arxiv.org/abs/hep-th/9606040} {arXiv:hep-th/9606040} \BibitemShut
  {NoStop}%
\bibitem [{\citenamefont {Padmanabhan}(1997)}]{Padmanabhan:1996ap}%
  \BibitemOpen
  \bibfield  {author} {\bibinfo {author} {\bibfnamefont {T.}~\bibnamefont
  {Padmanabhan}},\ }\bibfield  {title} {\enquote {\bibinfo {title} {{Duality
  and zero point length of space-time}},}\ }\href {\doibase
  10.1103/PhysRevLett.78.1854} {\bibfield  {journal} {\bibinfo  {journal}
  {Phys. Rev. Lett.}\ }\textbf {\bibinfo {volume} {78}},\ \bibinfo {pages}
  {1854--1857} (\bibinfo {year} {1997})},\ \Eprint
  {http://arxiv.org/abs/hep-th/9608182} {arXiv:hep-th/9608182} \BibitemShut
  {NoStop}%
\bibitem [{\citenamefont {Nicolini}\ \emph {et~al.}(2019)\citenamefont
  {Nicolini}, \citenamefont {Spallucci},\ and\ \citenamefont
  {Wondrak}}]{Nicolini:2019irw}%
  \BibitemOpen
  \bibfield  {author} {\bibinfo {author} {\bibfnamefont {Piero}\ \bibnamefont
  {Nicolini}}, \bibinfo {author} {\bibfnamefont {Euro}\ \bibnamefont
  {Spallucci}}, \ and\ \bibinfo {author} {\bibfnamefont {Michael~F.}\
  \bibnamefont {Wondrak}},\ }\bibfield  {title} {\enquote {\bibinfo {title}
  {{Quantum Corrected Black Holes from String T-Duality}},}\ }\href {\doibase
  10.1016/j.physletb.2019.134888} {\bibfield  {journal} {\bibinfo  {journal}
  {Phys. Lett. B}\ }\textbf {\bibinfo {volume} {797}},\ \bibinfo {pages}
  {134888} (\bibinfo {year} {2019})},\ \Eprint
  {http://arxiv.org/abs/1902.11242} {arXiv:1902.11242 [gr-qc]} \BibitemShut
  {NoStop}%
\bibitem [{\citenamefont {Gaete}\ \emph {et~al.}(2022)\citenamefont {Gaete},
  \citenamefont {Jusufi},\ and\ \citenamefont {Nicolini}}]{Gaete:2022ukm}%
  \BibitemOpen
  \bibfield  {author} {\bibinfo {author} {\bibfnamefont {Patricio}\
  \bibnamefont {Gaete}}, \bibinfo {author} {\bibfnamefont {Kimet}\ \bibnamefont
  {Jusufi}}, \ and\ \bibinfo {author} {\bibfnamefont {Piero}\ \bibnamefont
  {Nicolini}},\ }\bibfield  {title} {\enquote {\bibinfo {title} {{Charged black
  holes from T-duality}},}\ }\href {\doibase 10.1016/j.physletb.2022.137546}
  {\bibfield  {journal} {\bibinfo  {journal} {Phys. Lett. B}\ }\textbf
  {\bibinfo {volume} {835}},\ \bibinfo {pages} {137546} (\bibinfo {year}
  {2022})},\ \Eprint {http://arxiv.org/abs/2205.15441} {arXiv:2205.15441
  [hep-th]} \BibitemShut {NoStop}%
\bibitem [{\citenamefont {Nicolini}(2022)}]{Nicolini:2022rlz}%
  \BibitemOpen
  \bibfield  {author} {\bibinfo {author} {\bibfnamefont {Piero}\ \bibnamefont
  {Nicolini}},\ }\bibfield  {title} {\enquote {\bibinfo {title} {{Quantum
  gravity and the zero point length}},}\ }\href {\doibase
  10.1007/s10714-022-02995-4} {\bibfield  {journal} {\bibinfo  {journal} {Gen.
  Rel. Grav.}\ }\textbf {\bibinfo {volume} {54}},\ \bibinfo {pages} {106}
  (\bibinfo {year} {2022})},\ \Eprint {http://arxiv.org/abs/2208.05390}
  {arXiv:2208.05390 [hep-th]} \BibitemShut {NoStop}%
\bibitem [{\citenamefont {Jusufi}(2023)}]{Jusufi:2023pzt}%
  \BibitemOpen
  \bibfield  {author} {\bibinfo {author} {\bibfnamefont {Kimet}\ \bibnamefont
  {Jusufi}},\ }\bibfield  {title} {\enquote {\bibinfo {title} {{Avoidance of
  singularity during the gravitational collapse with string T-duality
  effects}},}\ }\href {\doibase 10.3390/universe9010041} {\bibfield  {journal}
  {\bibinfo  {journal} {Universe}\ }\textbf {\bibinfo {volume} {9}},\ \bibinfo
  {pages} {41} (\bibinfo {year} {2023})},\ \Eprint
  {http://arxiv.org/abs/2301.03590} {arXiv:2301.03590 [gr-qc]} \BibitemShut
  {NoStop}%
\bibitem [{\citenamefont {Jusufi}\ and\ \citenamefont
  {Sheykhi}(2023)}]{Jusufi:2022mir}%
  \BibitemOpen
  \bibfield  {author} {\bibinfo {author} {\bibfnamefont {Kimet}\ \bibnamefont
  {Jusufi}}\ and\ \bibinfo {author} {\bibfnamefont {Ahmad}\ \bibnamefont
  {Sheykhi}},\ }\bibfield  {title} {\enquote {\bibinfo {title} {{Entropic
  corrections to Friedmann equations and bouncing universe due to the
  zero-point length}},}\ }\href {\doibase 10.1016/j.physletb.2022.137621}
  {\bibfield  {journal} {\bibinfo  {journal} {Phys. Lett. B}\ }\textbf
  {\bibinfo {volume} {836}},\ \bibinfo {pages} {137621} (\bibinfo {year}
  {2023})},\ \Eprint {http://arxiv.org/abs/2210.01584} {arXiv:2210.01584
  [gr-qc]} \BibitemShut {NoStop}%
\bibitem [{\citenamefont {Millano}\ \emph {et~al.}(2023)\citenamefont
  {Millano}, \citenamefont {Jusufi},\ and\ \citenamefont
  {Leon}}]{Millano:2023ahb}%
  \BibitemOpen
  \bibfield  {author} {\bibinfo {author} {\bibfnamefont {Alfredo~D.}\
  \bibnamefont {Millano}}, \bibinfo {author} {\bibfnamefont {Kimet}\
  \bibnamefont {Jusufi}}, \ and\ \bibinfo {author} {\bibfnamefont {Genly}\
  \bibnamefont {Leon}},\ }\bibfield  {title} {\enquote {\bibinfo {title}
  {{Phase space analysis of the bouncing universe with stringy effects}},}\
  }\href@noop {} {\  (\bibinfo {year} {2023})},\ \Eprint
  {http://arxiv.org/abs/2302.00223} {arXiv:2302.00223 [gr-qc]} \BibitemShut
  {NoStop}%
\bibitem [{\citenamefont {Amati}\ \emph {et~al.}(1989)\citenamefont {Amati},
  \citenamefont {Ciafaloni},\ and\ \citenamefont {Veneziano}}]{Amati:1988tn}%
  \BibitemOpen
  \bibfield  {author} {\bibinfo {author} {\bibfnamefont {D.}~\bibnamefont
  {Amati}}, \bibinfo {author} {\bibfnamefont {M.}~\bibnamefont {Ciafaloni}}, \
  and\ \bibinfo {author} {\bibfnamefont {G.}~\bibnamefont {Veneziano}},\
  }\bibfield  {title} {\enquote {\bibinfo {title} {{Can Space-Time Be Probed
  Below the String Size?}}}\ }\href {\doibase 10.1016/0370-2693(89)91366-X}
  {\bibfield  {journal} {\bibinfo  {journal} {Phys. Lett. B}\ }\textbf
  {\bibinfo {volume} {216}},\ \bibinfo {pages} {41--47} (\bibinfo {year}
  {1989})}\BibitemShut {NoStop}%
\bibitem [{\citenamefont {Ali}\ \emph {et~al.}(2009)\citenamefont {Ali},
  \citenamefont {Das},\ and\ \citenamefont {Vagenas}}]{Ali:2009zq}%
  \BibitemOpen
  \bibfield  {author} {\bibinfo {author} {\bibfnamefont {Ahmed~Farag}\
  \bibnamefont {Ali}}, \bibinfo {author} {\bibfnamefont {Saurya}\ \bibnamefont
  {Das}}, \ and\ \bibinfo {author} {\bibfnamefont {Elias~C.}\ \bibnamefont
  {Vagenas}},\ }\bibfield  {title} {\enquote {\bibinfo {title} {{Discreteness
  of Space from the Generalized Uncertainty Principle}},}\ }\href {\doibase
  10.1016/j.physletb.2009.06.061} {\bibfield  {journal} {\bibinfo  {journal}
  {Phys. Lett. B}\ }\textbf {\bibinfo {volume} {678}},\ \bibinfo {pages}
  {497--499} (\bibinfo {year} {2009})},\ \Eprint
  {http://arxiv.org/abs/0906.5396} {arXiv:0906.5396 [hep-th]} \BibitemShut
  {NoStop}%
\bibitem [{\citenamefont {Kempf}\ \emph {et~al.}(1995)\citenamefont {Kempf},
  \citenamefont {Mangano},\ and\ \citenamefont {Mann}}]{Kempf:1994su}%
  \BibitemOpen
  \bibfield  {author} {\bibinfo {author} {\bibfnamefont {Achim}\ \bibnamefont
  {Kempf}}, \bibinfo {author} {\bibfnamefont {Gianpiero}\ \bibnamefont
  {Mangano}}, \ and\ \bibinfo {author} {\bibfnamefont {Robert~B.}\ \bibnamefont
  {Mann}},\ }\bibfield  {title} {\enquote {\bibinfo {title} {{Hilbert space
  representation of the minimal length uncertainty relation}},}\ }\href
  {\doibase 10.1103/PhysRevD.52.1108} {\bibfield  {journal} {\bibinfo
  {journal} {Phys. Rev. D}\ }\textbf {\bibinfo {volume} {52}},\ \bibinfo
  {pages} {1108--1118} (\bibinfo {year} {1995})},\ \Eprint
  {http://arxiv.org/abs/hep-th/9412167} {arXiv:hep-th/9412167} \BibitemShut
  {NoStop}%
\bibitem [{\citenamefont {Brau}(1999)}]{Brau:1999uv}%
  \BibitemOpen
  \bibfield  {author} {\bibinfo {author} {\bibfnamefont {F.}~\bibnamefont
  {Brau}},\ }\bibfield  {title} {\enquote {\bibinfo {title} {{Minimal length
  uncertainty relation and hydrogen atom}},}\ }\href {\doibase
  10.1088/0305-4470/32/44/308} {\bibfield  {journal} {\bibinfo  {journal} {J.
  Phys. A}\ }\textbf {\bibinfo {volume} {32}},\ \bibinfo {pages} {7691--7696}
  (\bibinfo {year} {1999})},\ \Eprint {http://arxiv.org/abs/quant-ph/9905033}
  {arXiv:quant-ph/9905033} \BibitemShut {NoStop}%
\bibitem [{\citenamefont {Maggiore}(1993)}]{Maggiore:1993rv}%
  \BibitemOpen
  \bibfield  {author} {\bibinfo {author} {\bibfnamefont {Michele}\ \bibnamefont
  {Maggiore}},\ }\bibfield  {title} {\enquote {\bibinfo {title} {{A Generalized
  uncertainty principle in quantum gravity}},}\ }\href {\doibase
  10.1016/0370-2693(93)91401-8} {\bibfield  {journal} {\bibinfo  {journal}
  {Phys. Lett. B}\ }\textbf {\bibinfo {volume} {304}},\ \bibinfo {pages}
  {65--69} (\bibinfo {year} {1993})},\ \Eprint
  {http://arxiv.org/abs/hep-th/9301067} {arXiv:hep-th/9301067} \BibitemShut
  {NoStop}%
\bibitem [{\citenamefont {Adler}(1994)}]{Adler:1993hm}%
  \BibitemOpen
  \bibfield  {author} {\bibinfo {author} {\bibfnamefont {Stephen~L.}\
  \bibnamefont {Adler}},\ }\bibfield  {title} {\enquote {\bibinfo {title}
  {{Generalized quantum dynamics}},}\ }\href {\doibase
  10.1016/0550-3213(94)90072-8} {\bibfield  {journal} {\bibinfo  {journal}
  {Nucl. Phys. B}\ }\textbf {\bibinfo {volume} {415}},\ \bibinfo {pages}
  {195--242} (\bibinfo {year} {1994})},\ \Eprint
  {http://arxiv.org/abs/hep-th/9306009} {arXiv:hep-th/9306009} \BibitemShut
  {NoStop}%
\bibitem [{\citenamefont {Scardigli}(1999)}]{Scardigli:1999jh}%
  \BibitemOpen
  \bibfield  {author} {\bibinfo {author} {\bibfnamefont {Fabio}\ \bibnamefont
  {Scardigli}},\ }\bibfield  {title} {\enquote {\bibinfo {title} {{Generalized
  uncertainty principle in quantum gravity from micro - black hole Gedanken
  experiment}},}\ }\href {\doibase 10.1016/S0370-2693(99)00167-7} {\bibfield
  {journal} {\bibinfo  {journal} {Phys. Lett. B}\ }\textbf {\bibinfo {volume}
  {452}},\ \bibinfo {pages} {39--44} (\bibinfo {year} {1999})},\ \Eprint
  {http://arxiv.org/abs/hep-th/9904025} {arXiv:hep-th/9904025} \BibitemShut
  {NoStop}%
\bibitem [{\citenamefont {Carr}\ \emph {et~al.}(2015)\citenamefont {Carr},
  \citenamefont {Mureika},\ and\ \citenamefont {Nicolini}}]{Carr:2015nqa}%
  \BibitemOpen
  \bibfield  {author} {\bibinfo {author} {\bibfnamefont {Bernard~J.}\
  \bibnamefont {Carr}}, \bibinfo {author} {\bibfnamefont {Jonas}\ \bibnamefont
  {Mureika}}, \ and\ \bibinfo {author} {\bibfnamefont {Piero}\ \bibnamefont
  {Nicolini}},\ }\bibfield  {title} {\enquote {\bibinfo {title} {{Sub-Planckian
  black holes and the Generalized Uncertainty Principle}},}\ }\href {\doibase
  10.1007/JHEP07(2015)052} {\bibfield  {journal} {\bibinfo  {journal} {JHEP}\
  }\textbf {\bibinfo {volume} {07}},\ \bibinfo {pages} {052} (\bibinfo {year}
  {2015})},\ \Eprint {http://arxiv.org/abs/1504.07637} {arXiv:1504.07637
  [gr-qc]} \BibitemShut {NoStop}%
\bibitem [{\citenamefont {Carr}(2022)}]{Carr:2022ndy}%
  \BibitemOpen
  \bibfield  {author} {\bibinfo {author} {\bibfnamefont {B.~J.}\ \bibnamefont
  {Carr}},\ }\bibfield  {title} {\enquote {\bibinfo {title} {{The Generalized
  Uncertainty Principle and higher dimensions: Linking black holes and
  elementary particles}},}\ }\href {\doibase 10.3389/fspas.2022.1008221}
  {\bibfield  {journal} {\bibinfo  {journal} {Front. Astron. Space Sci.}\
  }\textbf {\bibinfo {volume} {9}},\ \bibinfo {pages} {1008221} (\bibinfo
  {year} {2022})},\ \Eprint {http://arxiv.org/abs/2302.12609} {arXiv:2302.12609
  [gr-qc]} \BibitemShut {NoStop}%
\bibitem [{\citenamefont {Ali}\ \emph {et~al.}(2011)\citenamefont {Ali},
  \citenamefont {Das},\ and\ \citenamefont {Vagenas}}]{Ali:2011fa}%
  \BibitemOpen
  \bibfield  {author} {\bibinfo {author} {\bibfnamefont {Ahmed~Farag}\
  \bibnamefont {Ali}}, \bibinfo {author} {\bibfnamefont {Saurya}\ \bibnamefont
  {Das}}, \ and\ \bibinfo {author} {\bibfnamefont {Elias~C.}\ \bibnamefont
  {Vagenas}},\ }\bibfield  {title} {\enquote {\bibinfo {title} {{A proposal for
  testing Quantum Gravity in the lab}},}\ }\href {\doibase
  10.1103/PhysRevD.84.044013} {\bibfield  {journal} {\bibinfo  {journal} {Phys.
  Rev. D}\ }\textbf {\bibinfo {volume} {84}},\ \bibinfo {pages} {044013}
  (\bibinfo {year} {2011})},\ \Eprint {http://arxiv.org/abs/1107.3164}
  {arXiv:1107.3164 [hep-th]} \BibitemShut {NoStop}%
\bibitem [{\citenamefont {Das}\ and\ \citenamefont
  {Vagenas}(2008)}]{Das:2008kaa}%
  \BibitemOpen
  \bibfield  {author} {\bibinfo {author} {\bibfnamefont {Saurya}\ \bibnamefont
  {Das}}\ and\ \bibinfo {author} {\bibfnamefont {Elias~C.}\ \bibnamefont
  {Vagenas}},\ }\bibfield  {title} {\enquote {\bibinfo {title} {{Universality
  of Quantum Gravity Corrections}},}\ }\href {\doibase
  10.1103/PhysRevLett.101.221301} {\bibfield  {journal} {\bibinfo  {journal}
  {Phys. Rev. Lett.}\ }\textbf {\bibinfo {volume} {101}},\ \bibinfo {pages}
  {221301} (\bibinfo {year} {2008})},\ \Eprint {http://arxiv.org/abs/0810.5333}
  {arXiv:0810.5333 [hep-th]} \BibitemShut {NoStop}%
\bibitem [{\citenamefont {Pikovski}\ \emph {et~al.}(2012)\citenamefont
  {Pikovski}, \citenamefont {Vanner}, \citenamefont {Aspelmeyer}, \citenamefont
  {Kim},\ and\ \citenamefont {Brukner}}]{Pikovski:2011zk}%
  \BibitemOpen
  \bibfield  {author} {\bibinfo {author} {\bibfnamefont {Igor}\ \bibnamefont
  {Pikovski}}, \bibinfo {author} {\bibfnamefont {Michael~R.}\ \bibnamefont
  {Vanner}}, \bibinfo {author} {\bibfnamefont {Markus}\ \bibnamefont
  {Aspelmeyer}}, \bibinfo {author} {\bibfnamefont {M.~S.}\ \bibnamefont {Kim}},
  \ and\ \bibinfo {author} {\bibfnamefont {Caslav}\ \bibnamefont {Brukner}},\
  }\bibfield  {title} {\enquote {\bibinfo {title} {{Probing Planck-scale
  physics with quantum optics}},}\ }\href {\doibase 10.1038/nphys2262}
  {\bibfield  {journal} {\bibinfo  {journal} {Nature Phys.}\ }\textbf {\bibinfo
  {volume} {8}},\ \bibinfo {pages} {393--397} (\bibinfo {year} {2012})},\
  \Eprint {http://arxiv.org/abs/1111.1979} {arXiv:1111.1979 [quant-ph]}
  \BibitemShut {NoStop}%
\bibitem [{\citenamefont {Marin}\ \emph {et~al.}(2013)\citenamefont {Marin}
  \emph {et~al.}}]{Marin:2013pga}%
  \BibitemOpen
  \bibfield  {author} {\bibinfo {author} {\bibfnamefont {Francesco}\
  \bibnamefont {Marin}} \emph {et~al.},\ }\bibfield  {title} {\enquote
  {\bibinfo {title} {{Gravitational bar detectors set limits to Planck-scale
  physics on macroscopic variables}},}\ }\href {\doibase 10.1038/nphys2503}
  {\bibfield  {journal} {\bibinfo  {journal} {Nature Phys.}\ }\textbf {\bibinfo
  {volume} {9}},\ \bibinfo {pages} {71--73} (\bibinfo {year}
  {2013})}\BibitemShut {NoStop}%
\bibitem [{\citenamefont {Petruzziello}\ and\ \citenamefont
  {Illuminati}(2021)}]{Petruzziello:2020wkd}%
  \BibitemOpen
  \bibfield  {author} {\bibinfo {author} {\bibfnamefont {Luciano}\ \bibnamefont
  {Petruzziello}}\ and\ \bibinfo {author} {\bibfnamefont {Fabrizio}\
  \bibnamefont {Illuminati}},\ }\bibfield  {title} {\enquote {\bibinfo {title}
  {{Quantum gravitational decoherence from fluctuating minimal length and
  deformation parameter at the Planck scale}},}\ }\href {\doibase
  10.1038/s41467-021-24711-7} {\bibfield  {journal} {\bibinfo  {journal}
  {Nature Commun.}\ }\textbf {\bibinfo {volume} {12}},\ \bibinfo {pages} {4449}
  (\bibinfo {year} {2021})},\ \Eprint {http://arxiv.org/abs/2011.01255}
  {arXiv:2011.01255 [gr-qc]} \BibitemShut {NoStop}%
\bibitem [{\citenamefont {Feng}\ \emph {et~al.}(2017)\citenamefont {Feng},
  \citenamefont {Yang}, \citenamefont {Li},\ and\ \citenamefont
  {Zu}}]{Feng:2016tyt}%
  \BibitemOpen
  \bibfield  {author} {\bibinfo {author} {\bibfnamefont {Zhong-Wen}\
  \bibnamefont {Feng}}, \bibinfo {author} {\bibfnamefont {Shu-Zheng}\
  \bibnamefont {Yang}}, \bibinfo {author} {\bibfnamefont {Hui-Ling}\
  \bibnamefont {Li}}, \ and\ \bibinfo {author} {\bibfnamefont {Xiao-Tao}\
  \bibnamefont {Zu}},\ }\bibfield  {title} {\enquote {\bibinfo {title}
  {{Constraining the generalized uncertainty principle with the gravitational
  wave event GW150914}},}\ }\href {\doibase 10.1016/j.physletb.2017.02.043}
  {\bibfield  {journal} {\bibinfo  {journal} {Phys. Lett. B}\ }\textbf
  {\bibinfo {volume} {768}},\ \bibinfo {pages} {81--85} (\bibinfo {year}
  {2017})},\ \Eprint {http://arxiv.org/abs/1610.08549} {arXiv:1610.08549
  [hep-ph]} \BibitemShut {NoStop}%
\bibitem [{\citenamefont {Das}\ \emph {et~al.}(2021)\citenamefont {Das},
  \citenamefont {Das}, \citenamefont {Mansour},\ and\ \citenamefont
  {Vagenas}}]{das2021bounds}%
  \BibitemOpen
  \bibfield  {author} {\bibinfo {author} {\bibfnamefont {Ashmita}\ \bibnamefont
  {Das}}, \bibinfo {author} {\bibfnamefont {Saurya}\ \bibnamefont {Das}},
  \bibinfo {author} {\bibfnamefont {Noor~R}\ \bibnamefont {Mansour}}, \ and\
  \bibinfo {author} {\bibfnamefont {Elias~C}\ \bibnamefont {Vagenas}},\
  }\bibfield  {title} {\enquote {\bibinfo {title} {Bounds on gup parameters
  from gw150914 and gw190521},}\ }\href@noop {} {\bibfield  {journal} {\bibinfo
   {journal} {Physics Letters B}\ }\textbf {\bibinfo {volume} {819}},\ \bibinfo
  {pages} {136429} (\bibinfo {year} {2021})}\BibitemShut {NoStop}%
\bibitem [{\citenamefont {Moussa}\ \emph {et~al.}(2021)\citenamefont {Moussa},
  \citenamefont {Shababi},\ and\ \citenamefont {Farag~Ali}}]{Moussa:2021qlz}%
  \BibitemOpen
  \bibfield  {author} {\bibinfo {author} {\bibfnamefont {Mohamed}\ \bibnamefont
  {Moussa}}, \bibinfo {author} {\bibfnamefont {Homa}\ \bibnamefont {Shababi}},
  \ and\ \bibinfo {author} {\bibfnamefont {Ahmed}\ \bibnamefont {Farag~Ali}},\
  }\bibfield  {title} {\enquote {\bibinfo {title} {{Generalized uncertainty
  principle and stochastic gravitational wave background spectrum}},}\ }\href
  {\doibase 10.1016/j.physletb.2021.136071} {\bibfield  {journal} {\bibinfo
  {journal} {Phys. Lett. B}\ }\textbf {\bibinfo {volume} {814}},\ \bibinfo
  {pages} {136071} (\bibinfo {year} {2021})},\ \Eprint
  {http://arxiv.org/abs/2101.04747} {arXiv:2101.04747 [gr-qc]} \BibitemShut
  {NoStop}%
\bibitem [{\citenamefont {Scardigli}\ and\ \citenamefont
  {Casadio}(2015)}]{Scardigli:2014qka}%
  \BibitemOpen
  \bibfield  {author} {\bibinfo {author} {\bibfnamefont {Fabio}\ \bibnamefont
  {Scardigli}}\ and\ \bibinfo {author} {\bibfnamefont {Roberto}\ \bibnamefont
  {Casadio}},\ }\bibfield  {title} {\enquote {\bibinfo {title} {{Gravitational
  tests of the Generalized Uncertainty Principle}},}\ }\href {\doibase
  10.1140/epjc/s10052-015-3635-y} {\bibfield  {journal} {\bibinfo  {journal}
  {Eur. Phys. J. C}\ }\textbf {\bibinfo {volume} {75}},\ \bibinfo {pages} {425}
  (\bibinfo {year} {2015})},\ \Eprint {http://arxiv.org/abs/1407.0113}
  {arXiv:1407.0113 [hep-th]} \BibitemShut {NoStop}%
\bibitem [{\citenamefont {Hemeda}\ \emph {et~al.}(2024)\citenamefont {Hemeda},
  \citenamefont {Alshal}, \citenamefont {Ali},\ and\ \citenamefont
  {Vagenas}}]{Hemeda:2022dnd}%
  \BibitemOpen
  \bibfield  {author} {\bibinfo {author} {\bibfnamefont {Mohammed}\
  \bibnamefont {Hemeda}}, \bibinfo {author} {\bibfnamefont {Hassan}\
  \bibnamefont {Alshal}}, \bibinfo {author} {\bibfnamefont {Ahmed~Farag}\
  \bibnamefont {Ali}}, \ and\ \bibinfo {author} {\bibfnamefont {Elias~C.}\
  \bibnamefont {Vagenas}},\ }\bibfield  {title} {\enquote {\bibinfo {title}
  {{Gravitational observations and LQGUP}},}\ }\href {\doibase
  10.1016/j.nuclphysb.2024.116456} {\bibfield  {journal} {\bibinfo  {journal}
  {Nucl. Phys. B}\ }\textbf {\bibinfo {volume} {1000}},\ \bibinfo {pages}
  {116456} (\bibinfo {year} {2024})},\ \Eprint
  {http://arxiv.org/abs/2208.04686} {arXiv:2208.04686 [gr-qc]} \BibitemShut
  {NoStop}%
\bibitem [{\citenamefont {Kumar}\ and\ \citenamefont
  {Plenio}(2020)}]{Kumar:2019bnd}%
  \BibitemOpen
  \bibfield  {author} {\bibinfo {author} {\bibfnamefont {Shreya~P.}\
  \bibnamefont {Kumar}}\ and\ \bibinfo {author} {\bibfnamefont {Martin~B.}\
  \bibnamefont {Plenio}},\ }\bibfield  {title} {\enquote {\bibinfo {title} {{On
  Quantum Gravity Tests with Composite Particles}},}\ }\href {\doibase
  10.1038/s41467-020-17518-5} {\bibfield  {journal} {\bibinfo  {journal}
  {Nature Commun.}\ }\textbf {\bibinfo {volume} {11}},\ \bibinfo {pages} {3900}
  (\bibinfo {year} {2020})},\ \Eprint {http://arxiv.org/abs/1908.11164}
  {arXiv:1908.11164 [quant-ph]} \BibitemShut {NoStop}%
\bibitem [{\citenamefont {Moradpour}\ \emph {et~al.}(2019)\citenamefont
  {Moradpour}, \citenamefont {Ziaie}, \citenamefont {Ghaffari},\ and\
  \citenamefont {Feleppa}}]{Moradpour:2019wpj}%
  \BibitemOpen
  \bibfield  {author} {\bibinfo {author} {\bibfnamefont {H.}~\bibnamefont
  {Moradpour}}, \bibinfo {author} {\bibfnamefont {A.~H.}\ \bibnamefont
  {Ziaie}}, \bibinfo {author} {\bibfnamefont {S.}~\bibnamefont {Ghaffari}}, \
  and\ \bibinfo {author} {\bibfnamefont {F.}~\bibnamefont {Feleppa}},\
  }\bibfield  {title} {\enquote {\bibinfo {title} {{The generalized and
  extended uncertainty principles and their implications on the Jeans mass}},}\
  }\href {\doibase 10.1093/mnrasl/slz098} {\bibfield  {journal} {\bibinfo
  {journal} {Mon. Not. Roy. Astron. Soc.}\ }\textbf {\bibinfo {volume} {488}},\
  \bibinfo {pages} {L69--L74} (\bibinfo {year} {2019})},\ \Eprint
  {http://arxiv.org/abs/1907.12940} {arXiv:1907.12940 [gr-qc]} \BibitemShut
  {NoStop}%
\bibitem [{\citenamefont {Iorio}\ \emph {et~al.}(2018)\citenamefont {Iorio},
  \citenamefont {Pais}, \citenamefont {Elmashad}, \citenamefont {Ali},
  \citenamefont {Faizal},\ and\ \citenamefont {Abou-Salem}}]{Iorio:2017vtw}%
  \BibitemOpen
  \bibfield  {author} {\bibinfo {author} {\bibfnamefont {A.}~\bibnamefont
  {Iorio}}, \bibinfo {author} {\bibfnamefont {P.}~\bibnamefont {Pais}},
  \bibinfo {author} {\bibfnamefont {I.~A.}\ \bibnamefont {Elmashad}}, \bibinfo
  {author} {\bibfnamefont {A.~F.}\ \bibnamefont {Ali}}, \bibinfo {author}
  {\bibfnamefont {Mir}\ \bibnamefont {Faizal}}, \ and\ \bibinfo {author}
  {\bibfnamefont {L.~I.}\ \bibnamefont {Abou-Salem}},\ }\bibfield  {title}
  {\enquote {\bibinfo {title} {{Generalized Dirac structure beyond the linear
  regime in graphene}},}\ }\href {\doibase 10.1142/S0218271818500803}
  {\bibfield  {journal} {\bibinfo  {journal} {Int. J. Mod. Phys. D}\ }\textbf
  {\bibinfo {volume} {27}},\ \bibinfo {pages} {1850080} (\bibinfo {year}
  {2018})},\ \Eprint {http://arxiv.org/abs/1706.01332} {arXiv:1706.01332
  [physics.gen-ph]} \BibitemShut {NoStop}%
\bibitem [{\citenamefont {Gao}\ and\ \citenamefont
  {Zhan}(2016)}]{gao2016constraining}%
  \BibitemOpen
  \bibfield  {author} {\bibinfo {author} {\bibfnamefont {Dongfeng}\
  \bibnamefont {Gao}}\ and\ \bibinfo {author} {\bibfnamefont {Mingsheng}\
  \bibnamefont {Zhan}},\ }\bibfield  {title} {\enquote {\bibinfo {title}
  {Constraining the generalized uncertainty principle with cold atoms},}\
  }\href@noop {} {\bibfield  {journal} {\bibinfo  {journal} {Physical Review
  A}\ }\textbf {\bibinfo {volume} {94}},\ \bibinfo {pages} {013607} (\bibinfo
  {year} {2016})}\BibitemShut {NoStop}%
\bibitem [{\citenamefont {Bawaj}\ \emph {et~al.}(2015)\citenamefont {Bawaj}
  \emph {et~al.}}]{Bawaj:2014cda}%
  \BibitemOpen
  \bibfield  {author} {\bibinfo {author} {\bibfnamefont {Mateusz}\ \bibnamefont
  {Bawaj}} \emph {et~al.},\ }\bibfield  {title} {\enquote {\bibinfo {title}
  {{Probing deformed commutators with macroscopic harmonic oscillators}},}\
  }\href {\doibase 10.1038/ncomms8503} {\bibfield  {journal} {\bibinfo
  {journal} {Nature Commun.}\ }\textbf {\bibinfo {volume} {6}},\ \bibinfo
  {pages} {7503} (\bibinfo {year} {2015})},\ \Eprint
  {http://arxiv.org/abs/1411.6410} {arXiv:1411.6410 [gr-qc]} \BibitemShut
  {NoStop}%
\bibitem [{\citenamefont {Kober}(2010)}]{Kober:2010sj}%
  \BibitemOpen
  \bibfield  {author} {\bibinfo {author} {\bibfnamefont {Martin}\ \bibnamefont
  {Kober}},\ }\bibfield  {title} {\enquote {\bibinfo {title} {{Gauge Theories
  under Incorporation of a Generalized Uncertainty Principle}},}\ }\href
  {\doibase 10.1103/PhysRevD.82.085017} {\bibfield  {journal} {\bibinfo
  {journal} {Phys. Rev. D}\ }\textbf {\bibinfo {volume} {82}},\ \bibinfo
  {pages} {085017} (\bibinfo {year} {2010})},\ \Eprint
  {http://arxiv.org/abs/1008.0154} {arXiv:1008.0154 [physics.gen-ph]}
  \BibitemShut {NoStop}%
\bibitem [{\citenamefont {Girdhar}\ and\ \citenamefont
  {Doherty}(2020)}]{Girdhar:2020kfl}%
  \BibitemOpen
  \bibfield  {author} {\bibinfo {author} {\bibfnamefont {Parth}\ \bibnamefont
  {Girdhar}}\ and\ \bibinfo {author} {\bibfnamefont {Andrew~C.}\ \bibnamefont
  {Doherty}},\ }\bibfield  {title} {\enquote {\bibinfo {title} {{Testing
  generalised uncertainty principles through quantum noise}},}\ }\href
  {\doibase 10.1088/1367-2630/abb43c} {\bibfield  {journal} {\bibinfo
  {journal} {New J. Phys.}\ }\textbf {\bibinfo {volume} {22}},\ \bibinfo
  {pages} {093073} (\bibinfo {year} {2020})},\ \Eprint
  {http://arxiv.org/abs/2005.08984} {arXiv:2005.08984 [quant-ph]} \BibitemShut
  {NoStop}%
\bibitem [{\citenamefont {Segreto}\ and\ \citenamefont
  {Montani}(2022)}]{Segreto:2022xyv}%
  \BibitemOpen
  \bibfield  {author} {\bibinfo {author} {\bibfnamefont {Sebastiano}\
  \bibnamefont {Segreto}}\ and\ \bibinfo {author} {\bibfnamefont {Giovanni}\
  \bibnamefont {Montani}},\ }\bibfield  {title} {\enquote {\bibinfo {title}
  {{Extended GUP formulation with and without truncation in momentum space}},}\
  }\href@noop {} {\  (\bibinfo {year} {2022})},\ \Eprint
  {http://arxiv.org/abs/2208.03101} {arXiv:2208.03101 [quant-ph]} \BibitemShut
  {NoStop}%
\bibitem [{\citenamefont {Ali}\ and\ \citenamefont
  {Wojnar}(2024)}]{Ali:2024tbd}%
  \BibitemOpen
  \bibfield  {author} {\bibinfo {author} {\bibfnamefont {Ahmed~Farag}\
  \bibnamefont {Ali}}\ and\ \bibinfo {author} {\bibfnamefont {Aneta}\
  \bibnamefont {Wojnar}},\ }\bibfield  {title} {\enquote {\bibinfo {title} {{A
  covariant tapestry of linear GUP, metric-affine gravity, their Poincar\'e
  algebra and entropy bound}},}\ }\href {\doibase 10.1088/1361-6382/ad3ac7}
  {\bibfield  {journal} {\bibinfo  {journal} {Class. Quant. Grav.}\ }\textbf
  {\bibinfo {volume} {41}},\ \bibinfo {pages} {105001} (\bibinfo {year}
  {2024})},\ \Eprint {http://arxiv.org/abs/2401.05941} {arXiv:2401.05941
  [gr-qc]} \BibitemShut {NoStop}%
\bibitem [{\citenamefont {Ashoorioon}\ \emph {et~al.}(2005)\citenamefont
  {Ashoorioon}, \citenamefont {Kempf},\ and\ \citenamefont
  {Mann}}]{Ashoorioon:2004vm}%
  \BibitemOpen
  \bibfield  {author} {\bibinfo {author} {\bibfnamefont {A.}~\bibnamefont
  {Ashoorioon}}, \bibinfo {author} {\bibfnamefont {Achim}\ \bibnamefont
  {Kempf}}, \ and\ \bibinfo {author} {\bibfnamefont {Robert~B.}\ \bibnamefont
  {Mann}},\ }\bibfield  {title} {\enquote {\bibinfo {title} {{Minimum length
  cutoff in inflation and uniqueness of the action}},}\ }\href {\doibase
  10.1103/PhysRevD.71.023503} {\bibfield  {journal} {\bibinfo  {journal} {Phys.
  Rev. D}\ }\textbf {\bibinfo {volume} {71}},\ \bibinfo {pages} {023503}
  (\bibinfo {year} {2005})},\ \Eprint {http://arxiv.org/abs/astro-ph/0410139}
  {arXiv:astro-ph/0410139} \BibitemShut {NoStop}%
\bibitem [{\citenamefont {Easther}\ \emph {et~al.}(2003)\citenamefont
  {Easther}, \citenamefont {Greene}, \citenamefont {Kinney},\ and\
  \citenamefont {Shiu}}]{Easther:2001fz}%
  \BibitemOpen
  \bibfield  {author} {\bibinfo {author} {\bibfnamefont {Richard}\ \bibnamefont
  {Easther}}, \bibinfo {author} {\bibfnamefont {Brian~R.}\ \bibnamefont
  {Greene}}, \bibinfo {author} {\bibfnamefont {William~H.}\ \bibnamefont
  {Kinney}}, \ and\ \bibinfo {author} {\bibfnamefont {Gary}\ \bibnamefont
  {Shiu}},\ }\bibfield  {title} {\enquote {\bibinfo {title} {{Imprints of short
  distance physics on inflationary cosmology}},}\ }\href {\doibase
  10.1103/PhysRevD.67.063508} {\bibfield  {journal} {\bibinfo  {journal} {Phys.
  Rev. D}\ }\textbf {\bibinfo {volume} {67}},\ \bibinfo {pages} {063508}
  (\bibinfo {year} {2003})},\ \Eprint {http://arxiv.org/abs/hep-th/0110226}
  {arXiv:hep-th/0110226} \BibitemShut {NoStop}%
\bibitem [{\citenamefont {Dabrowski}\ and\ \citenamefont
  {Wagner}(2019)}]{dabrowski2019extended}%
  \BibitemOpen
  \bibfield  {author} {\bibinfo {author} {\bibfnamefont {Mariusz~P}\
  \bibnamefont {Dabrowski}}\ and\ \bibinfo {author} {\bibfnamefont {Fabian}\
  \bibnamefont {Wagner}},\ }\bibfield  {title} {\enquote {\bibinfo {title}
  {Extended uncertainty principle for rindler and cosmological horizons},}\
  }\href@noop {} {\bibfield  {journal} {\bibinfo  {journal} {The European
  Physical Journal C}\ }\textbf {\bibinfo {volume} {79}},\ \bibinfo {pages}
  {1--8} (\bibinfo {year} {2019})}\BibitemShut {NoStop}%
\bibitem [{\citenamefont {Ali}\ and\ \citenamefont
  {Majumder}(2014)}]{Ali:2014hma}%
  \BibitemOpen
  \bibfield  {author} {\bibinfo {author} {\bibfnamefont {Ahmed~Farag}\
  \bibnamefont {Ali}}\ and\ \bibinfo {author} {\bibfnamefont {Barun}\
  \bibnamefont {Majumder}},\ }\bibfield  {title} {\enquote {\bibinfo {title}
  {{Towards a Cosmology with Minimal Length and Maximal Energy}},}\ }\href
  {\doibase 10.1088/0264-9381/31/21/215007} {\bibfield  {journal} {\bibinfo
  {journal} {Class. Quant. Grav.}\ }\textbf {\bibinfo {volume} {31}},\ \bibinfo
  {pages} {215007} (\bibinfo {year} {2014})},\ \Eprint
  {http://arxiv.org/abs/1402.5104} {arXiv:1402.5104 [gr-qc]} \BibitemShut
  {NoStop}%
\bibitem [{\citenamefont {Easther}\ \emph {et~al.}(2001)\citenamefont
  {Easther}, \citenamefont {Greene}, \citenamefont {Kinney},\ and\
  \citenamefont {Shiu}}]{Easther:2001fi}%
  \BibitemOpen
  \bibfield  {author} {\bibinfo {author} {\bibfnamefont {Richard}\ \bibnamefont
  {Easther}}, \bibinfo {author} {\bibfnamefont {Brian~R.}\ \bibnamefont
  {Greene}}, \bibinfo {author} {\bibfnamefont {William~H.}\ \bibnamefont
  {Kinney}}, \ and\ \bibinfo {author} {\bibfnamefont {Gary}\ \bibnamefont
  {Shiu}},\ }\bibfield  {title} {\enquote {\bibinfo {title} {{Inflation as a
  probe of short distance physics}},}\ }\href {\doibase
  10.1103/PhysRevD.64.103502} {\bibfield  {journal} {\bibinfo  {journal} {Phys.
  Rev. D}\ }\textbf {\bibinfo {volume} {64}},\ \bibinfo {pages} {103502}
  (\bibinfo {year} {2001})},\ \Eprint {http://arxiv.org/abs/hep-th/0104102}
  {arXiv:hep-th/0104102} \BibitemShut {NoStop}%
\bibitem [{\citenamefont {Ali}\ \emph {et~al.}(2015)\citenamefont {Ali},
  \citenamefont {Faizal},\ and\ \citenamefont {Khalil}}]{Ali:2015ola}%
  \BibitemOpen
  \bibfield  {author} {\bibinfo {author} {\bibfnamefont {Ahmed~Farag}\
  \bibnamefont {Ali}}, \bibinfo {author} {\bibfnamefont {Mir}\ \bibnamefont
  {Faizal}}, \ and\ \bibinfo {author} {\bibfnamefont {Mohammed~M.}\
  \bibnamefont {Khalil}},\ }\bibfield  {title} {\enquote {\bibinfo {title}
  {{Short Distance Physics of the Inflationary de Sitter Universe}},}\ }\href
  {\doibase 10.1088/1475-7516/2015/9/025} {\bibfield  {journal} {\bibinfo
  {journal} {JCAP}\ }\textbf {\bibinfo {volume} {09}},\ \bibinfo {pages} {025}
  (\bibinfo {year} {2015})},\ \Eprint {http://arxiv.org/abs/1505.06963}
  {arXiv:1505.06963 [gr-qc]} \BibitemShut {NoStop}%
\bibitem [{\citenamefont {Scardigli}\ and\ \citenamefont
  {Casadio}(2003)}]{Scardigli:2003kr}%
  \BibitemOpen
  \bibfield  {author} {\bibinfo {author} {\bibfnamefont {Fabio}\ \bibnamefont
  {Scardigli}}\ and\ \bibinfo {author} {\bibfnamefont {Roberto}\ \bibnamefont
  {Casadio}},\ }\bibfield  {title} {\enquote {\bibinfo {title} {{Generalized
  uncertainty principle, extra dimensions and holography}},}\ }\href {\doibase
  10.1088/0264-9381/20/18/305} {\bibfield  {journal} {\bibinfo  {journal}
  {Class. Quant. Grav.}\ }\textbf {\bibinfo {volume} {20}},\ \bibinfo {pages}
  {3915--3926} (\bibinfo {year} {2003})},\ \Eprint
  {http://arxiv.org/abs/hep-th/0307174} {arXiv:hep-th/0307174} \BibitemShut
  {NoStop}%
\bibitem [{\citenamefont {Addazi}\ \emph {et~al.}(2021)\citenamefont {Addazi}
  \emph {et~al.}}]{Addazi:2021xuf}%
  \BibitemOpen
  \bibfield  {author} {\bibinfo {author} {\bibfnamefont {A.}~\bibnamefont
  {Addazi}} \emph {et~al.},\ }\bibfield  {title} {\enquote {\bibinfo {title}
  {{Quantum gravity phenomenology at the dawn of the multi-messenger era -- A
  review}},}\ }\href@noop {} {\  (\bibinfo {year} {2021})},\ \Eprint
  {http://arxiv.org/abs/2111.05659} {arXiv:2111.05659 [hep-ph]} \BibitemShut
  {NoStop}%
\bibitem [{\citenamefont {Hossenfelder}(2013)}]{Hossenfelder:2012jw}%
  \BibitemOpen
  \bibfield  {author} {\bibinfo {author} {\bibfnamefont {Sabine}\ \bibnamefont
  {Hossenfelder}},\ }\bibfield  {title} {\enquote {\bibinfo {title} {{Minimal
  Length Scale Scenarios for Quantum Gravity}},}\ }\href {\doibase
  10.12942/lrr-2013-2} {\bibfield  {journal} {\bibinfo  {journal} {Living Rev.
  Rel.}\ }\textbf {\bibinfo {volume} {16}},\ \bibinfo {pages} {2} (\bibinfo
  {year} {2013})},\ \Eprint {http://arxiv.org/abs/1203.6191} {arXiv:1203.6191
  [gr-qc]} \BibitemShut {NoStop}%
\bibitem [{\citenamefont {Bahrami}\ \emph {et~al.}(2014)\citenamefont
  {Bahrami}, \citenamefont {Gro\ss{}ardt}, \citenamefont {Donadi},\ and\
  \citenamefont {Bassi}}]{Bahrami:2014gwa}%
  \BibitemOpen
  \bibfield  {author} {\bibinfo {author} {\bibfnamefont {Mohammad}\
  \bibnamefont {Bahrami}}, \bibinfo {author} {\bibfnamefont {Andr\'e}\
  \bibnamefont {Gro\ss{}ardt}}, \bibinfo {author} {\bibfnamefont {Sandro}\
  \bibnamefont {Donadi}}, \ and\ \bibinfo {author} {\bibfnamefont {Angelo}\
  \bibnamefont {Bassi}},\ }\bibfield  {title} {\enquote {\bibinfo {title} {{The
  Schroedinger-Newton equation and its foundations}},}\ }\href {\doibase
  10.1088/1367-2630/16/11/115007} {\bibfield  {journal} {\bibinfo  {journal}
  {New J. Phys.}\ }\textbf {\bibinfo {volume} {16}},\ \bibinfo {pages} {115007}
  (\bibinfo {year} {2014})},\ \Eprint {http://arxiv.org/abs/1407.4370}
  {arXiv:1407.4370 [quant-ph]} \BibitemShut {NoStop}%
\bibitem [{\citenamefont {Gaete}\ and\ \citenamefont
  {Nicolini}(2022)}]{Gaete:2022une}%
  \BibitemOpen
  \bibfield  {author} {\bibinfo {author} {\bibfnamefont {Patricio}\
  \bibnamefont {Gaete}}\ and\ \bibinfo {author} {\bibfnamefont {Piero}\
  \bibnamefont {Nicolini}},\ }\bibfield  {title} {\enquote {\bibinfo {title}
  {{Finite electrodynamics from T-duality}},}\ }\href {\doibase
  10.1016/j.physletb.2022.137100} {\bibfield  {journal} {\bibinfo  {journal}
  {Phys. Lett. B}\ }\textbf {\bibinfo {volume} {829}},\ \bibinfo {pages}
  {137100} (\bibinfo {year} {2022})},\ \Eprint
  {http://arxiv.org/abs/2202.09311} {arXiv:2202.09311 [hep-th]} \BibitemShut
  {NoStop}%
\bibitem [{\citenamefont {Mureika}(2019)}]{Mureika:2018gxl}%
  \BibitemOpen
  \bibfield  {author} {\bibinfo {author} {\bibfnamefont {J.~R.}\ \bibnamefont
  {Mureika}},\ }\bibfield  {title} {\enquote {\bibinfo {title} {{Extended
  Uncertainty Principle Black Holes}},}\ }\href {\doibase
  10.1016/j.physletb.2018.12.009} {\bibfield  {journal} {\bibinfo  {journal}
  {Phys. Lett. B}\ }\textbf {\bibinfo {volume} {789}},\ \bibinfo {pages}
  {88--92} (\bibinfo {year} {2019})},\ \Eprint
  {http://arxiv.org/abs/1812.01999} {arXiv:1812.01999 [gr-qc]} \BibitemShut
  {NoStop}%
\bibitem [{\citenamefont {Magueijo}\ and\ \citenamefont
  {Smolin}(2002)}]{Magueijo:2001cr}%
  \BibitemOpen
  \bibfield  {author} {\bibinfo {author} {\bibfnamefont {Joao}\ \bibnamefont
  {Magueijo}}\ and\ \bibinfo {author} {\bibfnamefont {Lee}\ \bibnamefont
  {Smolin}},\ }\bibfield  {title} {\enquote {\bibinfo {title} {{Lorentz
  invariance with an invariant energy scale}},}\ }\href {\doibase
  10.1103/PhysRevLett.88.190403} {\bibfield  {journal} {\bibinfo  {journal}
  {Phys. Rev. Lett.}\ }\textbf {\bibinfo {volume} {88}},\ \bibinfo {pages}
  {190403} (\bibinfo {year} {2002})},\ \Eprint
  {http://arxiv.org/abs/hep-th/0112090} {arXiv:hep-th/0112090} \BibitemShut
  {NoStop}%
\bibitem [{\citenamefont {Magueijo}\ and\ \citenamefont
  {Smolin}(2003)}]{Magueijo:2002am}%
  \BibitemOpen
  \bibfield  {author} {\bibinfo {author} {\bibfnamefont {Joao}\ \bibnamefont
  {Magueijo}}\ and\ \bibinfo {author} {\bibfnamefont {Lee}\ \bibnamefont
  {Smolin}},\ }\bibfield  {title} {\enquote {\bibinfo {title} {{Generalized
  Lorentz invariance with an invariant energy scale}},}\ }\href {\doibase
  10.1103/PhysRevD.67.044017} {\bibfield  {journal} {\bibinfo  {journal} {Phys.
  Rev. D}\ }\textbf {\bibinfo {volume} {67}},\ \bibinfo {pages} {044017}
  (\bibinfo {year} {2003})},\ \Eprint {http://arxiv.org/abs/gr-qc/0207085}
  {arXiv:gr-qc/0207085} \BibitemShut {NoStop}%
\bibitem [{\citenamefont {Cortes}\ and\ \citenamefont
  {Gamboa}(2005)}]{Cortes:2004qn}%
  \BibitemOpen
  \bibfield  {author} {\bibinfo {author} {\bibfnamefont {J.~L.}\ \bibnamefont
  {Cortes}}\ and\ \bibinfo {author} {\bibfnamefont {J.}~\bibnamefont
  {Gamboa}},\ }\bibfield  {title} {\enquote {\bibinfo {title} {{Quantum
  uncertainty in doubly special relativity}},}\ }\href {\doibase
  10.1103/PhysRevD.71.065015} {\bibfield  {journal} {\bibinfo  {journal} {Phys.
  Rev. D}\ }\textbf {\bibinfo {volume} {71}},\ \bibinfo {pages} {065015}
  (\bibinfo {year} {2005})},\ \Eprint {http://arxiv.org/abs/hep-th/0405285}
  {arXiv:hep-th/0405285} \BibitemShut {NoStop}%
\bibitem [{\citenamefont {Ali}\ and\ \citenamefont
  {Majumder}(2021)}]{Ali:2021oml}%
  \BibitemOpen
  \bibfield  {author} {\bibinfo {author} {\bibfnamefont {Ahmed~Farag}\
  \bibnamefont {Ali}}\ and\ \bibinfo {author} {\bibfnamefont {Barun}\
  \bibnamefont {Majumder}},\ }\bibfield  {title} {\enquote {\bibinfo {title}
  {{Discreteness of space from anisotropic spin\textendash{}orbit
  interaction}},}\ }\href {\doibase 10.1140/epjc/s10052-021-09168-8} {\bibfield
   {journal} {\bibinfo  {journal} {Eur. Phys. J. C}\ }\textbf {\bibinfo
  {volume} {81}},\ \bibinfo {pages} {360} (\bibinfo {year} {2021})},\ \Eprint
  {http://arxiv.org/abs/2104.14563} {arXiv:2104.14563 [gr-qc]} \BibitemShut
  {NoStop}%
\bibitem [{\citenamefont {Ali}(2022)}]{Ali:2022jna}%
  \BibitemOpen
  \bibfield  {author} {\bibinfo {author} {\bibfnamefont {Ahmed~Farag}\
  \bibnamefont {Ali}},\ }\bibfield  {title} {\enquote {\bibinfo {title} {{EPR
  and Linear GUP: Completeness}},}\ }\href@noop {} {\  (\bibinfo {year}
  {2022})},\ \Eprint {http://arxiv.org/abs/2210.13974} {arXiv:2210.13974
  [quant-ph]} \BibitemShut {NoStop}%
\bibitem [{\citenamefont {Buoninfante}\ \emph {et~al.}(2022)\citenamefont
  {Buoninfante}, \citenamefont {Luciano}, \citenamefont {Petruzziello},\ and\
  \citenamefont {Scardigli}}]{Buoninfante:2020guu}%
  \BibitemOpen
  \bibfield  {author} {\bibinfo {author} {\bibfnamefont {Luca}\ \bibnamefont
  {Buoninfante}}, \bibinfo {author} {\bibfnamefont {Giuseppe~Gaetano}\
  \bibnamefont {Luciano}}, \bibinfo {author} {\bibfnamefont {Luciano}\
  \bibnamefont {Petruzziello}}, \ and\ \bibinfo {author} {\bibfnamefont
  {Fabio}\ \bibnamefont {Scardigli}},\ }\bibfield  {title} {\enquote {\bibinfo
  {title} {{Bekenstein bound and uncertainty relations}},}\ }\href {\doibase
  10.1016/j.physletb.2021.136818} {\bibfield  {journal} {\bibinfo  {journal}
  {Phys. Lett. B}\ }\textbf {\bibinfo {volume} {824}},\ \bibinfo {pages}
  {136818} (\bibinfo {year} {2022})},\ \Eprint
  {http://arxiv.org/abs/2009.12530} {arXiv:2009.12530 [hep-th]} \BibitemShut
  {NoStop}%
\end{thebibliography}%

\end{document}